%% file: sig2015_R1.tex
\documentclass{sig-alternate-2013}

\usepackage{amssymb,amsfonts,amsmath,amscd,dsfont,mathrsfs}
\usepackage{graphicx,float,psfrag,epsfig}
\usepackage{wrapfig}
\usepackage{epstopdf,color}
\usepackage{hyperref,bbm}
\usepackage{balance}
\usepackage{caption}

\usepackage{algorithm}
\usepackage[noend]{algpseudocode}
\DeclareMathAlphabet{\mathpzc}{OT1}{pzc}{m}{it}

\newtheorem{propo}{Proposition}[section]

\newtheorem{theorem}[propo]{Theorem}
\newtheorem{remark}[propo]{Remark}

\newcommand{\reals}{{\mathbb R}}
\newcommand{\prob}{\mathbb P}
\newcommand{\E}{\mathbb E}
\newcommand{\V}{V}
\newcommand{\G}{G}

\newcommand{\hv}{\hat{v}}

\floatname{algorithm}{Protocol}

\makeatletter
\def\BState{\State\hskip-\ALG@thistlm}
\makeatother

\newfont{\mycrnotice}{ptmr8t at 7pt}
\newfont{\myconfname}{ptmri8t at 7pt}
\let\crnotice\mycrnotice%
\let\confname\myconfname%

\permission{Permission to make digital or hard copies of all or part of this work for personal or classroom use is granted without fee provided that copies are not made or distributed for profit or commercial advantage and that copies bear this notice and the full citation on the first page. Copyrights for components of this work owned by others than ACM must be honored. Abstracting with credit is permitted. To copy otherwise, or republish, to post on servers or to redistribute to lists, requires prior specific permission and/or a fee. Request permissions from permissions@acm.org.}
\conferenceinfo{SIGMETRICS'15,}{June 15--19, 2015, Portland, OR, USA.}
\copyrightetc{Copyright \copyright~2015 ACM \the\acmcopyr}
\crdata{978-1-4503-3486-0/15/06\ ...\$15.00.\\
http://dx.doi.org/10.1145/2745844.2745866}

\begin{document}

\title{Spy vs. Spy: Rumor Source Obfuscation}


\numberofauthors{4}
\author{
\alignauthor Giulia Fanti \\
\affaddr{University of California, Berkeley} \\
\email{fanti@berkeley.edu}
\alignauthor Peter Kairouz\\
\affaddr{University of Illinois at Urbana-Champaign} \\
\email{kairouz2@illinois.edu}
\alignauthor Sewoong Oh\\
\affaddr{University of Illinois at Urbana-Champaign} \\
\email{swoh@illinois.edu}
\and
\alignauthor Pramod Viswanath\\
\affaddr{University of Illinois at Urbana-Champaign} \\
\email{pramodv@illinois.edu}
}



\date{ }
\maketitle

\maketitle

\begin{abstract}
Anonymous messaging platforms, such as Secret, Yik Yak and Whisper,  have emerged as important social media for
sharing one's thoughts without the fear of being judged by friends, family, or the public.
Further, such anonymous platforms are crucial in nations with authoritarian governments;
the right to free expression and sometimes the personal safety of the author of the message  depend on anonymity.
Whether for fear of judgment or personal endangerment,
it is crucial to keep anonymous the identity of the user who initially posted
a sensitive message.
In this paper, we consider an adversary who observes
a snapshot of the spread of a message at a certain time.
Recent advances in rumor source detection shows that the existing  messaging protocols
are  vulnerable against such an adversary.
We introduce a novel messaging protocol, which we call adaptive diffusion,
and show that it spreads the messages fast and achieves a perfect obfuscation of the source when the underlying contact network is an
infinite regular tree:
all users with the message are nearly equally likely to have been the origin of the message.
Experiments on a sampled Facebook network show that it
effectively hides the location of the source even when the graph is finite, irregular and has cycles.
\end{abstract}
%
%
%
%

%
%


\input{intro_v5}

\input{example_v5}

\input{main_v5}

\input{general_v4}


\input{discussion}

\bibliographystyle{plain}
\bibliography{sigmetrics2015}

\appendix

\input{protocol}

\input{proof}
\end{document}

%% file: intro_v5.tex
\section{Introduction}
\label{sec:intro}



Microblogging platforms form a core aspect of the fabric of the present Internet; popular examples include Twitter and Facebook. 
Users propagate short messages (texts, images, videos) through the platform via local friendship links.
The forwarding of messages often occurs through built-in mechanisms that rely on user input, such as clicking ``like" or ``share" with regards to a particular post.
Brevity of message, fluidity of user interface, and trusted party communication combine to make these microblogging platforms a major communication mode of modern times.
There has been tremendous recent interest in the privacy implications of these platforms, as evidenced by the explosive growth of {\em anonymous microblogging} platforms, like Secret \cite{secret}, Whisper \cite{whisper} and Yik Yak \cite{yikyak}.
These platforms enable users to share messages with friends without leaking the message author's identity.
In such applications, it is crucial to hide the identity of the user who initially posted the message.

Existing anonymous messaging services
store both messages and authorship information on centralized servers, which makes them vulnerable to
government subpoenas, hacking, or direct company access.
A more robust solution would be to store this information in a distributed fashion;
each node would know only its own friends, and message authorship information would never be transmitted to any party.
Distributed systems are  more robust to monitoring due to lack of central points of failure.
However, even under distributed architectures,
simple anonymous messaging protocols (such as those used by commercial anonymous microblogging apps) are still vulnerable against an adversary with side information, as proved in recent advances in rumor source detection.
In this work, we study in depth a basic building block of the messaging protocol
 that would underpin truly anonymous microblogging platform 
-- {\em broadcasting a single message on a contact network with the goal of obfuscating the source} under strong adversarial conditions.
Specifically, we consider contact networks that represent a social graph among users of the service.

A natural strategy for preventing source identification by an adversary would be to spread the message as fast as possible; with reliable connection to infrastructure like the Internet, this could be in principle done nearly instantaneously.
If all users receive the message instantaneously, any user is equally likely to have been the source.
However, this strategy is not available in many of the key real-life scenarios we are considering.
For instance, in social networks,
messages are spread based on users approving the message via liking, sharing or retweeting (to enable social filtering and also to avoid spamming) -- this scenario naturally has inherent  random delays associated with when the user happens to encounter the message and whether or not she decides to ``like" the message. Indeed, standard models of rumor spreading in networks  explicitly model such random delays via a {\em diffusion} process: messages are spread independently over different edges with a fixed probability of spreading (discrete time model) or an exponential time to spread (continuous time model).

\noindent{\bf Related work.}
Anonymous communication has been a popular research topic for decades. 
For instance, anonymous {point-to-point} communication
allows a sender to communicate with a receiver without the receiver learning the sender's identity.
A great deal of successful work has emerged in this area, including Tor \cite{tor}, Freenet \cite{freenet}, Free Haven \cite{freeHavenProject}, and Tarzan \cite{tarzan}.
In contrast to this body of work, we address the problem of anonymously \emph{broadcasting} a message over
an underlying contact network (e.g., a social network).
Anonymous broadcast communication has been most studied in context of the dining cryptographers' (DC) problem.
We diverge from the vast literature on this topic \cite{chaum88,corrigan2010dissent,goel2003herbivore,golle2004dining,von2003k} in approach and formulation.
We consider statistical spreading models rather than cryptographic encodings,
accommodate  computationally unbounded adversaries,  and consider arbitrary network structures rather than a fully connected network.


Within the realm of statistical message spreading models, the problem of detecting the origin of an epidemic or the source of a rumor
has been studied under the {\em diffusion} model.
Recent advances in \cite{SZ11a,SZ11b,WDZT14,PVF12,FC12,luo2013identify,ZY13,Austin1,Austin2,Austin3,Austin4} 
show that
it is possible to identify the source within a few hops with high probability.
Drawing an analogy to epidemics, we refer to a person who has received the message as `infected'
and the act of passing the message as `spreading the infection'.
Consider an adversary who has access to the underlying {\em contact network} of friendship links
and the snapshot of infected nodes at a certain time. The  problem of
locating a rumor source,   first posed in \cite{SZ11a}, naturally corresponds to
  {graph-centrality}-based inference algorithms: for a continuous time model,  \cite{SZ11a,SZ11b} used the
{rumor centrality} measure to correctly identify the source after time $T$ (with  probability converging to a positive number for large $d$-regular and random trees, and with probability proportional to $1/\sqrt{T}$ for lines).
The probability of identifying the source increases even further when multiple infections
from the same source are observed \cite{WDZT14}.
With multiple sources of infections,
spectral methods have been proposed for
estimating the number of sources and the set of source nodes in \cite{PVF12,FC12}.
When infected nodes are allowed to recover as in the susceptible-infected-recovered (SIR) model,
{Jordan centrality} was proposed in \cite{luo2013identify,ZY13} to estimate the source.
In \cite{ZY13}, it is shown that the Jordan center is still within a bounded hop distance from the
true source with high probability, independent of the number of infected nodes.
Under natural and diffusion-based message spreading -- as seen in almost every content-sharing platform today --
an adversary with some side-information can identify the rumor source with high confidence.
We overcome this vulnerability by asking the {reverse question}: can we {design} messaging protocols that spread fast while protecting the anonymity of the source?

\noindent{\bf Model.}
We focus on
anonymous 
microblogging
built atop an underlying contact network, such as Secret \cite{secret} over the network of phone contacts and Facebook friends.
In such systems, the designer has some control over the spreading rate,
by introducing artificial delays on top of the usual random delays due to users' approval of the messages. 
We model this physical setup as a discrete-time system, where
any individual receiving a message approves it immediately at the next timestep,
at which point the protocol determines
how much delay to introduce
before sending the message to each of her uninfected neighbors.
Given this control, the system designer wishes to design a spreading protocol that
makes inference on the source of the message difficult. 
The assumption that all nodes are willing to approve and pass the message is not new.
Such assumptions are common in the analysis of  rumor spreading \cite{SZ11a,SZ11b,ZY13},
and our deviation from those standard models is that we are operating in discrete time and approvals are immediate.

\noindent{\bf Adversary.}
Following the adversarial model assumed in prior work on rumor source detection \cite{SZ11a,SZ11b,ZY13},
we assume the adversary knows the whole underlying contact network and,
at a certain time, it observes a snapshot of the state of all the nodes, i.e. who has received the message thus far.
This adversary is strong in the sense that it sees the whole contact network
as well as every node's state, but it is also limited in the sense that
the adversary is not aware of when (or from whom) a particular node received the message.
This model captures an adversary that is able to indirectly observe the contents of users' devices without actively compromising the devices; for instance, if the message in question contains the time and location of a protest, then the adversary learns a snapshot of the infection at a given point in time by observing who attends the protest.
This adversarial model also captures an adversary that is able to monitor the network state more closely, but only at a high cost. As such, it cannot afford to continuously monitor state.
We design a new anonymous messaging protocol, which we call {\em adaptive diffusion},
that is inherently distributed and provides strong anonymity guarantees under this adversarial model.
We discuss other plausible adversarial models in Section \ref{sec:discussion}.

\noindent{\bf Spreading.}
At time $t=0$, a single user $v^*\in \V$ starts to spread a message on a contact network $\G=(\V,E)$ where
users and contacts are represented by nodes and edges, respectively.
Upon receiving the message, a node can send the message to
any of its neighbors.
We assume a discrete-time system and model the delays due to user approval and intermittent network access
via a deterministic delay of one time unit.
Therefore, a message always propagates with a delay of at least one time unit. Our goal is to introduce appropriate random delays
into the system in order to obfuscate the identity of the source $v^*$.
After $T$ timesteps, let $\V_T\subseteq \V$, $G_T$, and $N_T \triangleq |\V_T|$ denote the set of infected nodes, the subgraph of $G$ containing only $V_T$, and the number of infected nodes, respectively.
At a certain time $T$, an adversary observes the infected subgraph $G_T$
and produces an estimate $\hat{v}$ of the source $v^*$ of the message (with probability of
detection $P_D = \prob(\hat{v} = v^*)$). Since the adversary is assumed to not
have  any prior information on which node is likely to be the source,
we use the maximum likelihood estimator
\begin{equation}
\label{eq:maxlhest}
\hv_{\rm ML} = \arg\max_{v\in G_T} \prob(G_T|v).
\end{equation}
 We wish to achieve the following performance metrics.
\begin{itemize}
	\item [$(a)$] We say a protocol has an {\em order-optimal rate of spread} if the
	expected time for the message to reach $n$ nodes scales linearly compared to the time required by the fastest spreading protocol.
	\item [$(b)$] We say a protocol achieves a {\em perfect obfuscation} if the probability of source detection for the maximum likelihood estimator conditioned on $n$ nodes being infected is bounded by
	\begin{eqnarray}
		\prob\big(\,\hv_{\rm ML}=v^* | N_T=n\,\big) &=& \frac{1}{n}  + o\Big(\frac{1}{n}\Big)\;.
	\end{eqnarray}
\end{itemize}

\begin{figure}[tb]
	\vspace{-0.1cm}
	\begin{center}
	\includegraphics[width=.5\textwidth]{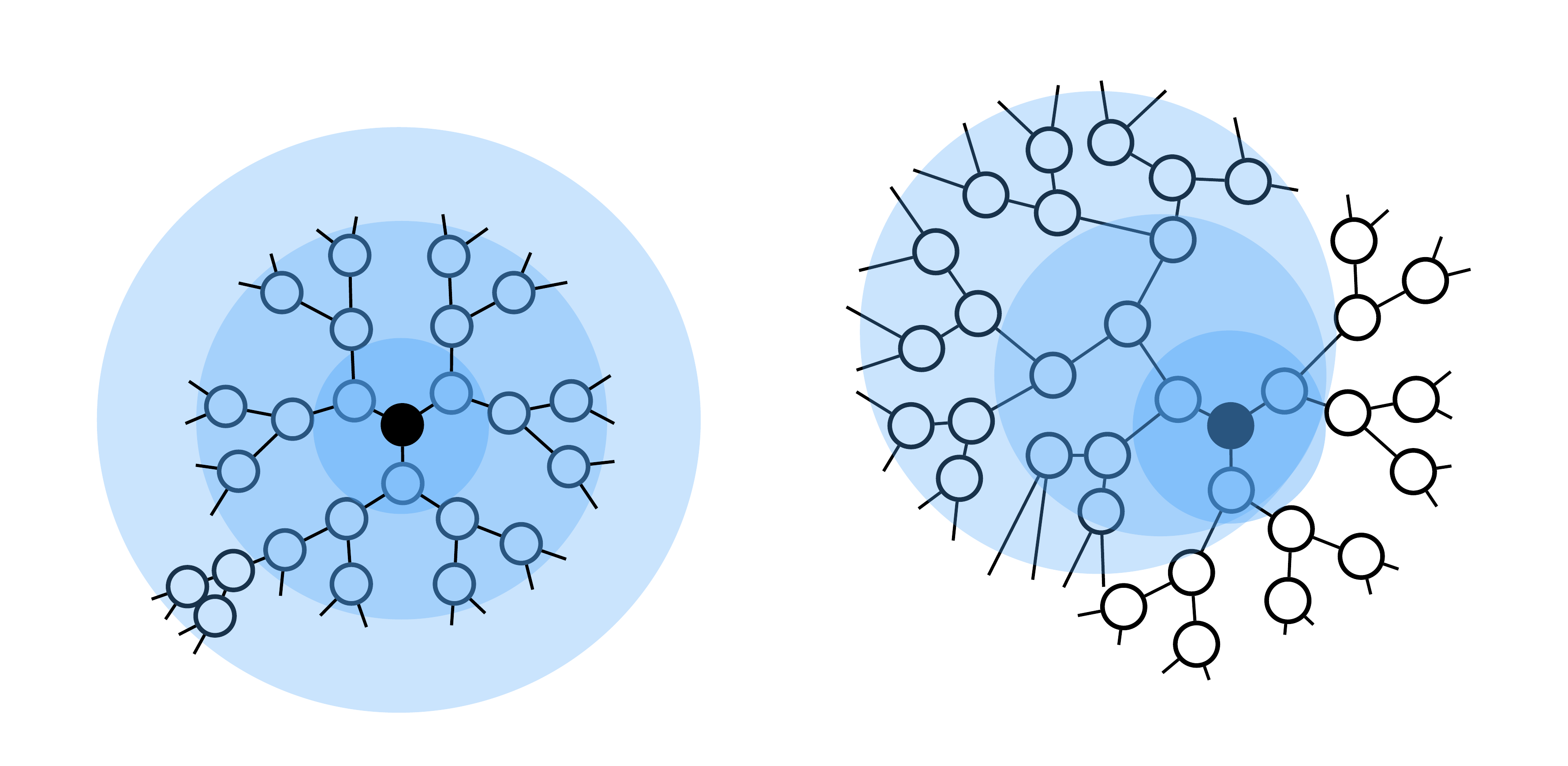}
	\end{center}
	\vspace{-.2cm}
	\caption{Illustration of a spread of infection when spreading immediately (left) and
	under the adaptive diffusion (right). }
	\label{fig:intro}
\end{figure}

\noindent{\bf Key insights.}
Figure \ref{fig:intro} (left) illustrates an example of the spread when the message is propagated immediately upon reception.
The source is indicated by a solid circle.
This scheme spreads the message fast but the source is trivially identified
as the center of the infected subgraph if the contact network is an infinite tree.
This is true independent of the infection size.
Even if we introduce some randomness at each node,
the source will still be identified within a few hops.
This is due to the fact that the source is close to some notion of the center of the infected subgraph  \cite{SZ11a,ZY13}.

Since we do not know \emph{a priori} when the adversary is going to attack,
the main challenge is to ensure that the source is equally likely to be anywhere in the infection at {\em any given time}.  Figure \ref{fig:intro} (right) illustrates the main idea of our approach:
we intentionally break the symmetry around the source.
This is achieved by combining two insights illustrated in the two warm-up examples in Section \ref{sec:examples}.
The first insight is that nodes farther away from the source should spread the infection faster.
The second insight is that the spread should be coordinated in order to maintain a symmetric structure
centered around a `virtual source' node. This leads to the source node being anywhere in the infected subgraph with equal probability.

\noindent{\bf Contributions.}
We introduce a novel messaging protocol, which we call {\em adaptive diffusion}, with provable author anonymity guarantees
against strong adversaries. 
Our protocol is inherently distributed and spreads messages fast, i.e.,
the time it takes adaptive diffusion to reach $n$ users is at most twice the time it takes the
fastest spreading scheme which immediately passes the message to all its neighbors.

We further prove that adaptive diffusion provides perfect obfuscation of the source
when applied to regular tree contact networks.
The source hides perfectly within all infected users, i.e.,
the likelihood of an infected user being the source of the infection is equal among all infected users.
For a more general class of graphs which can be finite, irregular and have cycles,
we provide results of numerical experiments on real-world social networks and synthetic networks
showing that the protocol hides the source at nearly the best possible level of obfuscation.



\noindent{\bf Organization.}
The remainder of this paper is organized as follows.
To warm up, we introduce, in Section \ref{sec:examples}, two messaging protocols customized for lines and trees.
Combining the key insights of these two approaches,
we introduce, in Section \ref{sec:adap_diff}, a new messaging protocol called {\em adaptive diffusion}
and analyze its performance theoretically and empirically, in Section \ref{sec:generalcn}.
Section \ref{sec:discussion} discusses limitations and future work.

%% file: example_v5.tex
\section{Warm-up Examples}
\label{sec:examples}

In this section, we discuss two special contact networks as warm-up examples:
a line and a regular tree with degree larger than two.
We provide two fully-distributed, customized messaging protocols, one for each case,
and show that these protocols
spread messages quickly while effectively hiding the source.
However, these protocols fail to protect the identity of the source
when applied to a broader class of contact networks.
In particular, the ``Line Protocol", which is developed for line contact networks,
reveals the source with high probability when applied to a regular tree with degree larger than two.
Similarly, the ``Tree Protocol", developed for regular tree contact networks with degree greater than two, reveals the source with high probability when
applied to a line.
In Section \ref{sec:adap_diff}, we introduce a novel messaging protocol, which we call
{\em adaptive diffusion}, that combines the key ideas behind the two approaches presented in this section.

\subsection{Spreading on a line}
\label{subsec:line}

Given a contact network of an infinite line,
consider the following deterministic spreading protocol. At time $t=1$,
the source node infects its left and right neighbors. At $t \geq 2$,
the leftmost and rightmost infected nodes spread the message to their uninfected neighbors.
Thus,  the message spreads one hop to the left and one hop to the right of the true source at each timestep.
This scheme spreads as fast as possible, infecting $N_T = 2T+1$ nodes at time $T$, but
the source is trivially identified as the center of the infection.

Adding a little bit of randomness can significantly decrease the probability of detection.
Consider a discrete-time {\em random diffusion} model with a parameter $p\in (0,1)$ where
at each time $t$, an infected node infects its uninfected neighbor with probability $p$.
Using the analysis from  \cite{SZ11a} where the continuous time version of
this protocol was studied,
we can show that this protocol spreads fast, infecting $\E[N_T] = 2pT+1$ nodes on average at time $T$. Further, the probability of source detection $P_D = \prob(\hv_{\rm ML}=v^*)$ for the maximum likelihood estimator scales as $1/\sqrt{p(1-p)T}$.
With $p=1/2$ for example, this gives a simple messaging protocol with a probability of source detection vanishing at a rate of $1/\sqrt{T}$.

\begin{figure}[h]
    \centering
  \includegraphics[width=.45\textwidth]{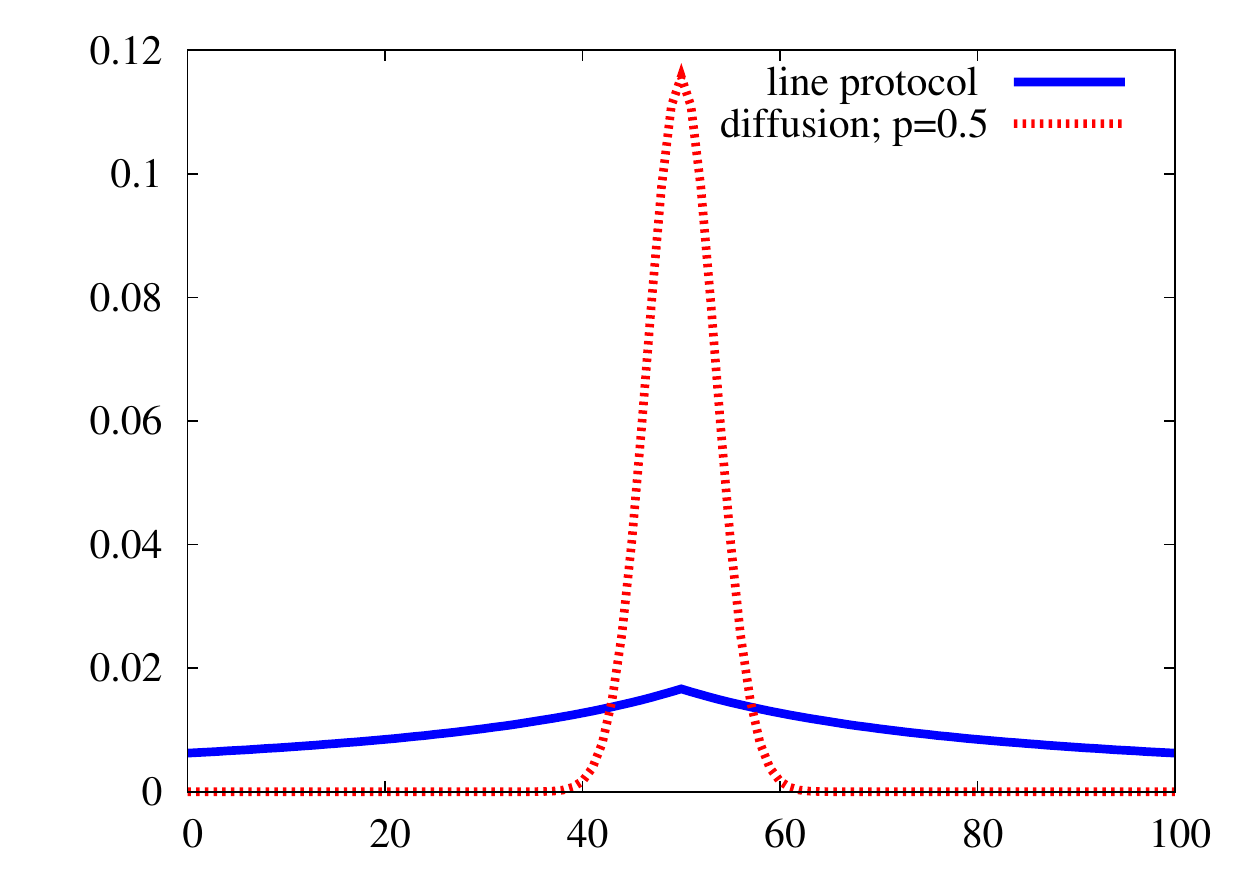}
  \put(-120,-5){Node $k$}
  \put(-225,55){\rotatebox{90}{\large $\prob(v^*=k|G_T)$}}
  \caption{The Line Protocol has a close-to-uniform posterior distribution
  $\prob(v^*=k\,|\, V_T=\{1,\ldots,101\})$. 
   }
  \label{fig:line}
\end{figure}

In what follows, we show that with an appropriate choice of {\em time-dependent} randomness,
we can achieve almost perfect source obfuscation without sacrificing
the spreading rate.
The key insight is to add randomness
such that all the infected nodes are (almost) equally likely to have been the origin of the infection (see Figure \ref{fig:line} and Equation \eqref{eq:lineposterior0}).
This can be achieved by adaptively choosing the spreading rate
such that {\em the farther away the infection is from the source the more likely it is to spread}.
We now apply this insight to design precisely how fast the spread should be for each infected node at any timestep.
A node $v$ is designed to infect a neighbor at time $t\in\{1,2,\ldots\}$ with probability
\begin{eqnarray}
	p_{v,t} &\triangleq&  \frac{\delta_H(v,v^*) + 1}{t+1} \;,
	\label{eq:linespread}
\end{eqnarray}
where $\delta_H(v,v^*)$ is the hop distance between an infected node $v$ at the boundary of infection and the source $v^*$. The details of this spreading model are summarized in Protocol \ref{alg:line} (Line Protocol).

The next proposition shows that this
protocol achieves the two main goals of an anonymous messaging protocol:
order-optimal spreading rate and close-to-perfect obfuscation.
\begin{propo}
\label{pro:line}
Suppose that the underlying contact network $G$ is an infinite line,
and one node $v^*$ in $G$ starts to spread a message according to
Protocol  \ref{alg:line} (Line Protocol) at time $t=0$.
At a certain time $T\geq0$ an adversary estimates the location of the source $v^*$
using the maximum likelihood estimator $\hv_{\rm ML}$ defined in Equation \eqref{eq:maxlhest}.
The following properties hold for the Line Protocol: 
\begin{itemize}
	\item[$(a)$] the expected number of infected nodes at time $T$ is $\mathbb{E}[N_T] = T+1$;
%
	
	\item[$(b)$] \vspace{-0.2cm}  the probability of source detection
	at time $T$ is upper bounded by
	\begin{eqnarray}
     \prob (\hv_{\rm ML}=v^* ) &\leq& \frac{2T+1}{(T+1)^2}
		\; ;  \text{ and }\label{eq:linedetection2}
	\end{eqnarray}
	
	\item[$(c)$] \vspace{-0.2cm} the expected hop-distance between the true source $v^*$ and its estimate $\hat{v}_{\rm ML}$  is lower bounded by
	\begin{eqnarray}
\E[\delta_H(v^*,\hat{v}_{\rm ML})] &\geq& \frac{T^3}{9(T+1)^2} \;.
		\label{eq:linedist}
	\end{eqnarray}
\end{itemize}
\end{propo}




The proof of the above proposition can be found in Appendix \ref{apndx:proof1}.
Compared to the (fastest-spreading) deterministic spreading model with a spreading rate of $N_T=2T+1$,
The Line Protocol is slower by a factor of $2$.
This type of constant-factor loss in the spreading rate is inevitable: the only way to deviate from the deterministic spreading model is
to introduce appropriate delays.
The probability of detection is $2/\E[N_T] + o(1/\E[N_T])$, which is almost perfect obfuscation up to a factor of $2$.
Further, the expected distance of the true source from the ML source estimate scales linearly with the size of the infection $\E[N_T]$, which is the best separation one can hope to achieve.

%

To illustrate the power of the Line Protocol, we consider a fixed $T$ and a finite ring graph of size larger than $2T+1$,
and compare the protocol to a simple random diffusion.
If the source $v^*$ is chosen uniformly at random on the ring and its message is spread according to the Line Protocol, then
the probability of the source being detected given a set of infected nodes $V_T$ is
	\begin{eqnarray}
		\prob\big(v^*=k \,\big|\, V_T\big) &=& \frac{1}{|V_T|} + O\Big(\frac{1}{|V_T|^2}\Big)\;,
		\label{eq:lineposterior0}
	\end{eqnarray}
for all $k\in V_T$ and $|V_T|\leq 2T+1$.
This follows from the exact computation of the posterior distribution, which can be found at the end of Appendix \ref{apndx:proof1}.
For an example with $|V_T|=101$,
Figure \ref{fig:line} illustrates how the Line Protocol flattens
the posterior distribution compared to the random diffusion model.
When messages are sent according to the random diffusion model,
the source can only hide in the central part, which has width $O(\sqrt{T})$,
leading to a probability of source detection on the order of $1/\sqrt{T}$ \cite{SZ11a}.

\begin{figure}
\begin{center}
	\includegraphics[width=.45\textwidth]{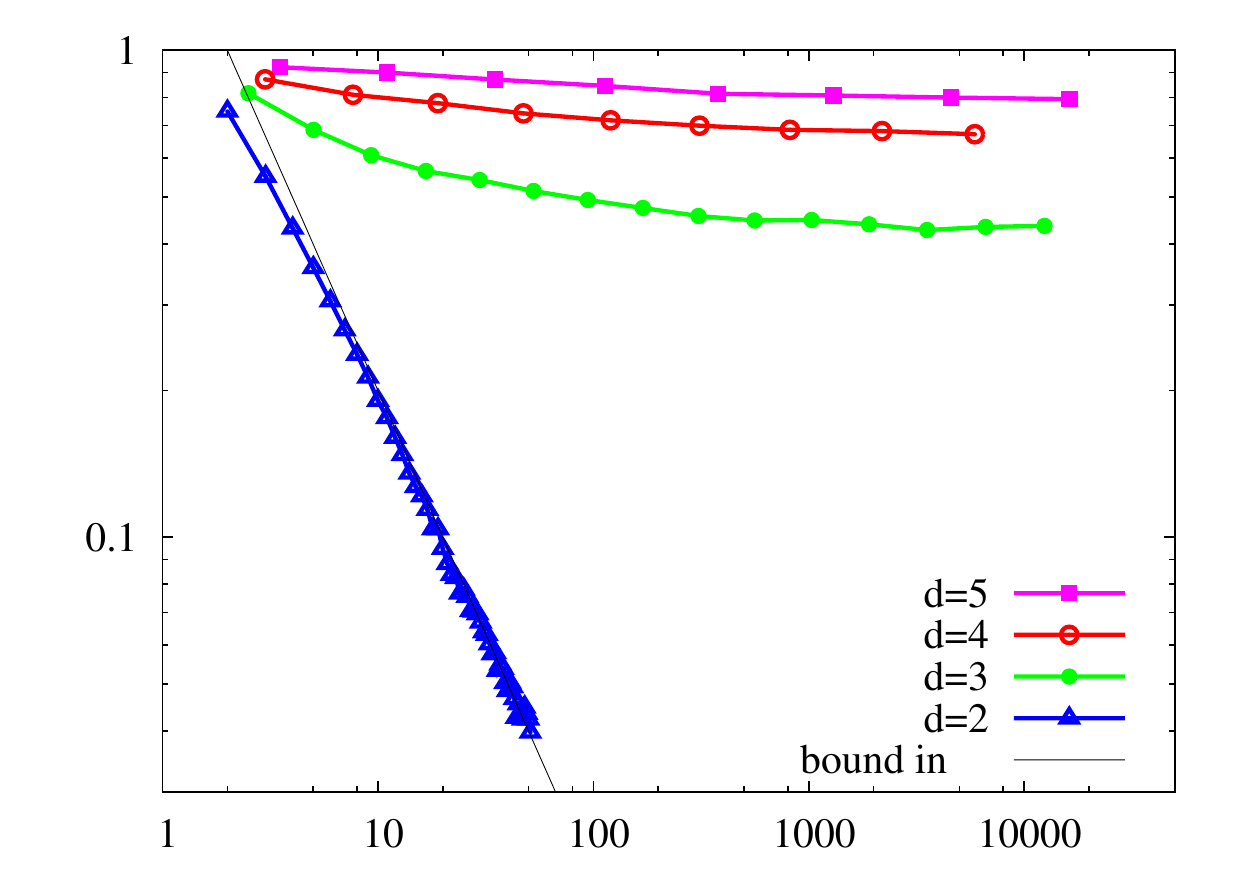}
	\put(-170,-5){Expected size of infection $\E[N_T]$}
	\put(-225,55){\rotatebox{90}{\large $\prob(\hv=v^*)$}}
	\put(-55,19){\tiny{ \eqref{eq:linedetection2}}}
  \caption{Detection probability versus the average size of infection on regular trees using Jordan center estimator, which is the ML estimate in case of a line ($d=2$). For $d>2$, the source is easily detected. }
  \label{fig:line2}
\end{center}
\end{figure}

On an infinite line, the Line Protocol provides maximum protection,
since the probability of detection scales as $1/\E[N_T]$ for any $T$.
When the Line Protocol is applied to regular trees with degree larger than two, the infected subgraph contains exponentially many paths  starting at $v^*$ of length close to $T$.
In such cases,  the Jordan center  (i.e., the node with the smallest maximum distance to every other node in the graph) matches the source with positive   probability, as shown in Figure \ref{fig:line2} for different $d$-regular trees.

 \subsection{Spreading on a regular tree}
\label{subsec:tree}

Consider the case when the underlying contact network is an infinite $d$-regular tree with $d$ larger than two.
Analogous to the line network, the fastest spreading protocol infects all the
uninfected neighbors at each timestep. This spreads fast, infecting $N_T=1+d((d-1)^T-1)/(d-2)$ nodes at time $T$,
but the source is trivially identified as the center
of the infected subtree.
In this case, the infected subtree is a
balanced regular tree where all leaves are at equal depth from the source.

Now consider a random diffusion model. At each timestep, each uninfected neighbor of an infected node
is independently infected with probability $p$.
In this case, $\E[N_T] = 1+ pd((d-1)^T-1)/(d-2)$, and it was shown in \cite{SZ11a}
that the probability of correct detection for the maximum likelihood estimator of the rumor source
is $ \prob(\hv_{\rm ML}=v^*) \geq C_d $ for some positive constant $C_d$ that only depends on the degree $d$. Hence, the source is only hidden in a constant number of nodes close to the center, even when
the total number of infected nodes is arbitrarily large.

\begin{figure}
    \centering
  \includegraphics[scale=0.45]{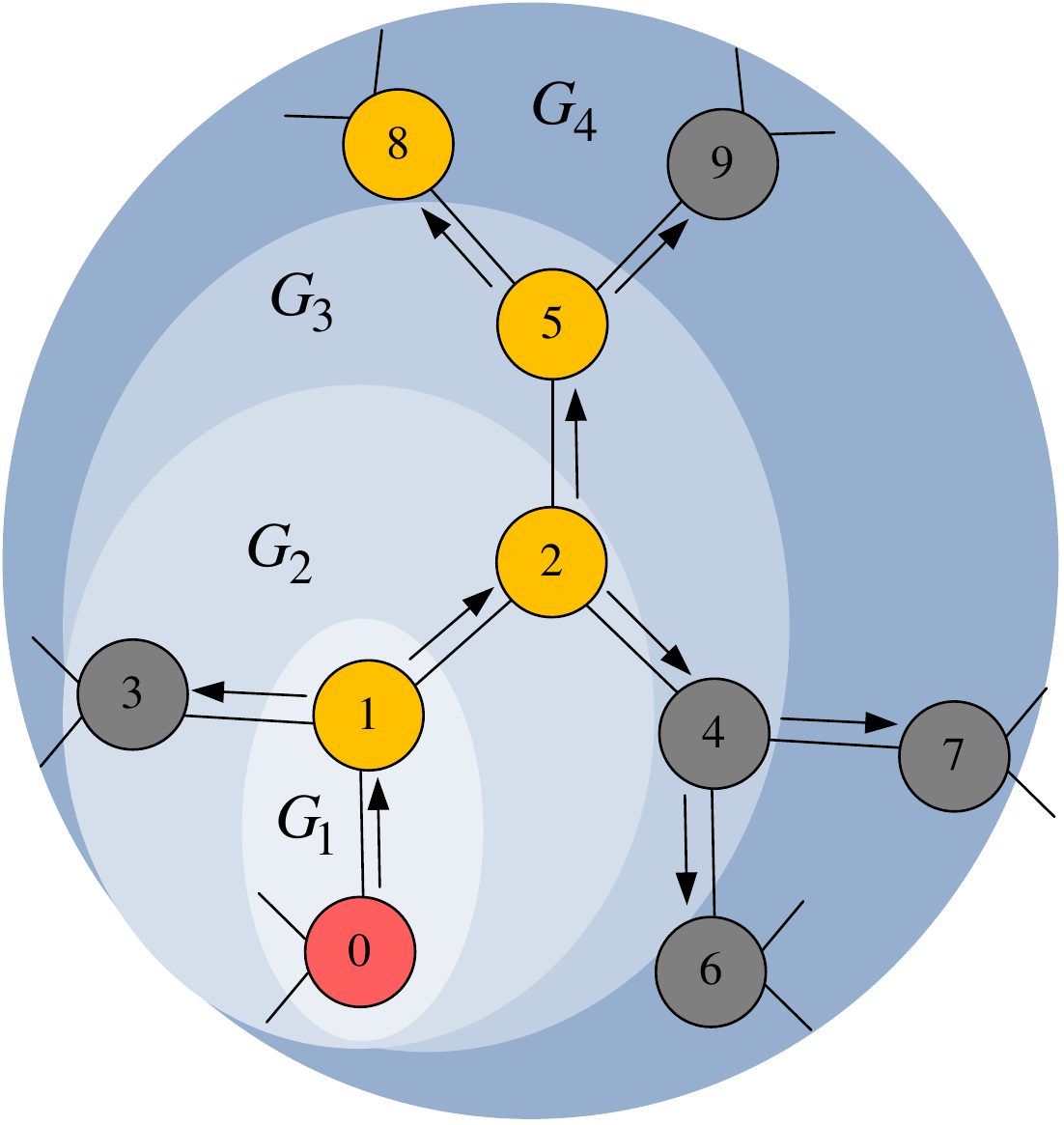}
  \caption{Spreading on a tree. The red node is the message source. Yellow nodes denote nodes that have been, are, or will be the center of the infected subtree.
}
  \label{fig:tree}
\end{figure}

We now present a protocol that spreads the message fast ($N_T=O((d-1)^{T/2})$)
and hides the source within a constant fraction of the infected nodes ($\prob(\hv_{\rm ML}=v^*)=O(1/N_T)$). This protocol
keeps the infected subtree balanced: at any time $t$, all the leaves of the infected subtree are at the same hop distance from its center. Further, as we will see next, the leaves of the infected subtree are equally likely to have been the source. Figure \ref{fig:tree} illustrates how this protocol spreads a message on a regular tree of degree 3. At $t=1$, node 0 (the message author) infects one of its neighbors (node 1 in this example) uniformly at random. Node 1 will be referred to as the virtual source at $t=1$. The virtual source at time $t$ is the center of the infected subtree at time $2t$.
At $t=2$, node $1$ infects all its uninfected neighbors, making the infected subgraph $G_2$ a balanced tree with node $1$ at the center.
Among the uninfected neighbors of node $1$,
one node is chosen to be the new virtual source (node $2$ in the example).
The message then spreads to the uninfected neighbors of node $2$ at time $t=3$,
and then to their neighbors at time $t=4$ making $G_4$ a balanced tree with node $2$ at the center. Notice that any given time $t$, all leaves are equally likely to have been the source. This follows form the symmetric structure of $G_t$.
The distributed implementation of this spreading algorithm is given in Protocol \ref{alg:tree} (Tree Protocol).

The Tree Protocol ensures that the source can hide among the leaf nodes of the infected subtree, i.e. all leaves are equally likely to have been the source. Since a significant fraction of the infected nodes are at the leaf,
this protocol achieves an almost {\em perfect obfuscation}.
\begin{propo}
\label{pro:tree}
Suppose that the underlying contact network $G$ is an infinite $d$-regular tree with $d>2$,
and one node $v^*$ in $G$ starts to spread a message according to
Protocol  \ref{alg:tree} (Tree Protocol) at time $t=0$.
At a certain time $T\geq1$ an adversary estimates the location of the source $v^*$
using the maximum likelihood estimator $\hv_{\rm ML}$.
Then the following properties hold for the Tree Protocol: 
\begin{itemize}
	\item[$(a)$] the number of infected nodes at time $T\geq1$ is at least
	\begin{equation}
	\label{eq:n_tree}
	N_T \;\geq \;
	\frac{(d-1)^{(T+1)/2}}{d-2}  \; ;
	\end{equation}
	
	\item[$(b)$] \vspace{-0.2cm} the probability of source detection for the maximum likelihood estimator at time $T$ is
	\begin{eqnarray}
\prob\big( \hv_{\rm ML} =v^* \big) &=& \frac{d-1}{2+(d-2)N_T} \; ; \text{ and }
\label{eq:pd_tree}
	\end{eqnarray}
	
	\item[$(c)$] \vspace{-0.2cm} the expected hop-distance between the true source $v^*$ and its estimate $\hat{v}$
	is lower bounded by
	\begin{eqnarray}
\E[\delta_H(v^*,\hv_{\rm ML})] &\geq& \frac{T}{2} \;.
		\label{eq:treedist}
	\end{eqnarray}
	\end{itemize}
\end{propo}

The proof of the above proposition can be found in Appendix \ref{apndx:proof2}.
Equation \eqref{eq:n_tree} shows that
the spreading rate of the Tree Protocol is $O((d-1)^{T/2})$, which is slower than
the deterministic spreading model that infects $O((d-1)^T)$ nodes at time $T$. This is inevitable, as we explained in relation to
the Line Protocol.

Although this protocol spreads fast and provides an almost perfect obfuscation on a tree with degree larger than two,  it fails 
when the contact network is a line.
There are only two leaves in a line, so at any given time $T$, the source can be detected with probability $1/2$,
independent of the size of the infected subgraph.
Another drawback of this approach, is that
even in the long run, not every node receives the message.
For instance, the neighbors of the source node that are not chosen in the first step are never infected.
In the following section, we address these issues and propose a new messaging protocol that combines the key ideas of both spreading models presented in this section.

%% file: main_v5.tex
\section{Adaptive Diffusion}
\label{sec:adap_diff}

Section \ref{sec:examples} showed that by changing the infection rate and direction based on state variables, the source can hide from the adversary. In particular, the messaging protocols presented in Sections \ref{subsec:line} and \ref{subsec:tree}
provide provable anonymity guarantees for line graphs and $d$-regular trees with $d>2$, respectively. However, the Line Protocol fails to protect the source on graphs with larger degree. Similarly, the Tree Protocol fails to protect the source on a line and does not pass the message to some of the nodes.
To overcome these challenges, we use ideas from the Line Protocol (nodes farther away from the source spread message faster)
and from the Tree Protocol (keep the infected subgraph balanced and keep the source closer to the leaves)
to design a protocol that achieves perfect obfuscation and spreads fast on all regular trees, including lines. We call this protocol \textit{adaptive diffusion} to emphasize the fact that unlike diffusion, the protocol adapts the infection rate and direction as a function of time.

\begin{figure}[b!]
    \centering
  \includegraphics[scale=0.5]{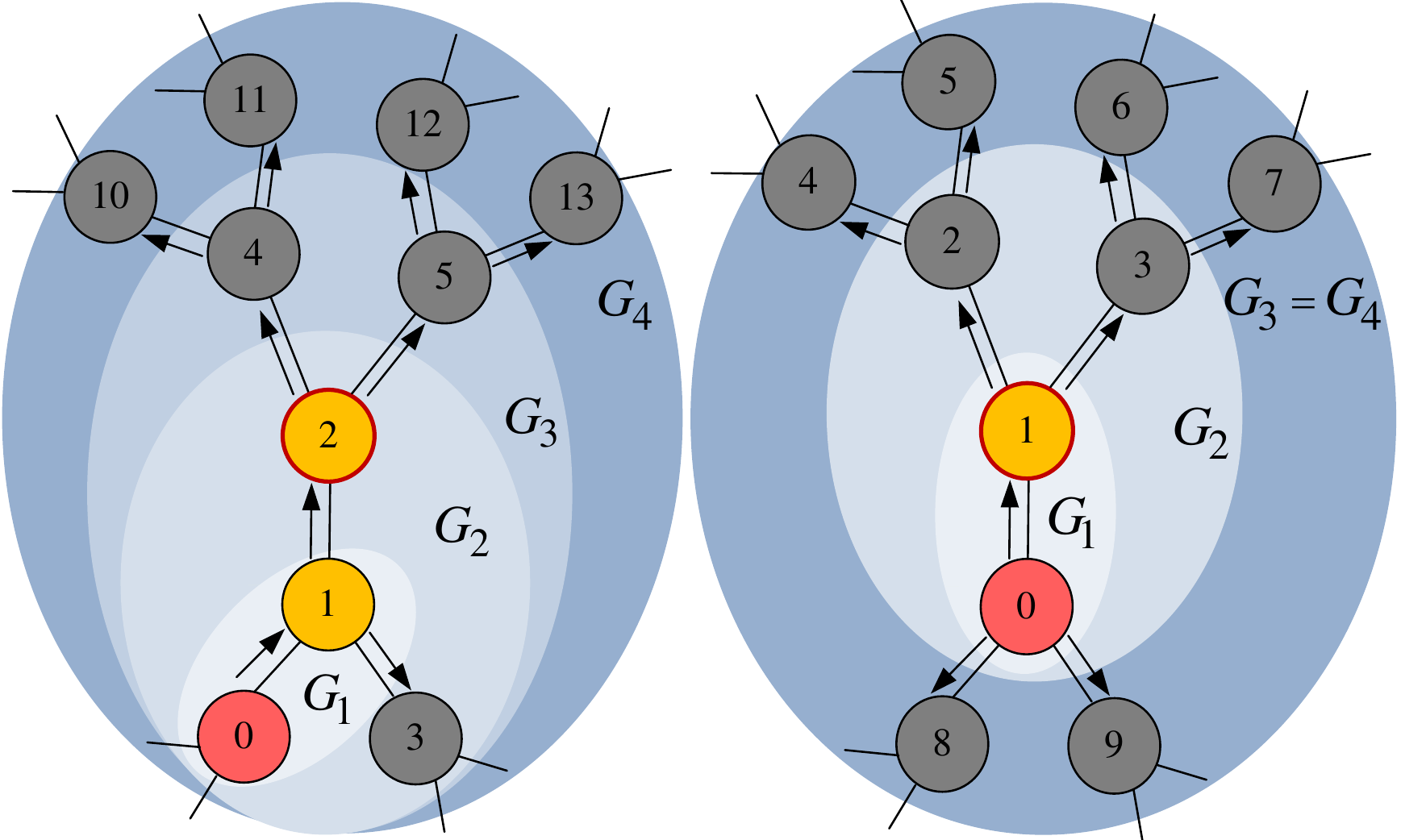}
  \caption{Adaptive diffusion over regular trees. Yellow nodes indicate the set of virtual sources (past and present), and for $T=4$, the virtual source node is outlined in red.}
  \label{fig:graph}
\end{figure}

We step through the intuition of the adaptive diffusion spreading model with an example, partially illustrated in Figure \ref{fig:graph}. Suppose that the underlying contact network is an infinite $d$-regular tree. As illustrated in Figure \ref{fig:graph}, we ensure that the infected subgraph $G_t$ at any even timestep $t\in\{2,4,\ldots\}$ is
a balanced tree of depth $t/2$, i.e. the hop distance from any leaf to the root (or the center of the graph) is $t/2$.
We call the root node of $G_t$ the ``virtual source'' at time $t$, and denote it by $v_t$.
We use $v_0=v^*$ to denote the true source.
To keep the regular structure at even timesteps,
we use the odd timesteps to transition from one regular subtree $G_t$ to another one $G_{t+2}$ with depth incremented by one.

Figure \ref{fig:graph} illustrates two sample evolutions of infection, as per adaptive diffusion.
The source $v^*=0$ starts the infection at $t=0$. At time $t=1$, node $0$ infects node $1$ and passes the virtual source token to it, i.e.  $v_2=1$ (we only define virtual sources for even timesteps). At time $t=1$, node $1$ infects its uninfected neighbors, nodes $2$ and $3$.
Notice that it requires two timesteps to infect nodes $\{1,2,3\}$ in order to spread infection to $G_2$, which is a balanced tree of depth $1$ rooted at node $v_2=1$.
At time $t=3$, the adaptive diffusion protocol has two choices, either to pass the virtual source token to one of node $1$'s neighbors
that is not a previous virtual source, for example node $2$ (Figure \ref{fig:graph} left),
or to keep the virtual source at node $1$ (Figure \ref{fig:graph} right).
In the former case, it again takes two timesteps to spread infection to $G_4$, which is a balanced tree of depth $2$ rooted at node $v_4=2$.
In the latter case, only one timestep is required, but we add a unit time delay to be consistent with the previous case. Hence, $G_3=G_4$ which is a balanced tree of depth $2$ rooted at node $v_4=1$.
This random infection process can be defined as a time-inhomogeneous (time-dependent) Markov chain over the state defined by the location of the current virtual source $\{v_t\}_{t\in\{0,2,4,\ldots\}}$.

By the symmetry of the underlying contact network (which we assume is an infinite $d$-regular tree) and
the fact that the next virtual source is chosen uniformly at random among the neighbors of the current virtual source,
it is sufficient to consider a Markov chain over the hop distance between the true source $v^*$ and $v_t$, the virtual source at time $t$.
Therefore, we design a Markov chain over the state
\begin{eqnarray*}
	h_t &=& \delta_H(v^*,v_t)\;,
\end{eqnarray*}
for even $t$.
Figure \ref{fig:graph} shows an example with $(h_2,h_4)= (1,2)$ on the left and
 $(h_2,h_4)=(1,1)$ on the right.

At every even timestep, the protocol randomly determines  whether
to keep the virtual source token ($h_{t+2} = h_t$) or to pass it ($h_{t+2}=h_t + 1$).
Using ideas from Section \ref{subsec:line},
we will construct an time-inhomogeneous Markov chain over $\{h_t\}_{t\in\{2,4,6,\ldots\}}$ by choosing appropriate transition probabilities as a function of
 time $t$ and current state $h_t$.
 For an even $t$, we denote this probability by
\begin{eqnarray}
	\alpha_d(t,h) &\triangleq& \prob\big(\,h_{t+2}=h_t | h_t=h\,\big) \;,
\end{eqnarray}
where the subscript $d$ denotes the degree of the underlying contact network.
For the running example, 
at $t=2$, the virtual source remains at the current node (right) with probability $\alpha_3(2,1)$,
or passes the virtual source to a neighbor with probability $1-\alpha_3(2,1)$ (left).
The parameters $\alpha_d(t,h)$ fully describe the transition probability of the Markov chain defined over $h_t \in \{1,2,\ldots,t/2\}$.
Let $p^{(t)} = [p^{(t)}_h]_{h\in\{1,\ldots,t/2\}}$
denote the distribution of the state of the Markov chain at time $t$, i.e.
$p^{(t)}_h = \prob(h_t = h)$.
The state transition can be represented as  the following $((t/2)+1)\times (t/2)$ dimensional column stochastic matrices:
\begin{eqnarray*}
	\label{eq:MC}
	p^{(t+2)} = \begin{bmatrix}
	\alpha_d(t,1) & & & \\
	1-\alpha_d(t,1) &\alpha_d(t,2) & & \\
	& 1-\alpha_d(t,2)& \ddots& \\
	& & \ddots& \alpha_d(t,t/2) \\
	& & &  1-\alpha_d(t,t/2) \\
	\end{bmatrix} p^{(t)}\;.
\end{eqnarray*}
We treat $h_t$ as strictly positive, because at time $t=0$, when $h_0=0$, the virtual source is always passed. Thus, $h_t\geq1$ afterwards. We design the the parameters $\alpha_d(t,h)$
to achieve perfect hiding. Precisely,
at all even $t$, we desire $p^{(t)}$ to be
\begin{eqnarray}
	\label{eq:MCp}
	p^{(t)} = \frac{d-2}{(d-1)^{t/2}-1}\begin{bmatrix}
		1\\
		(d-1)\\
		\vdots\\
		(d-1)^{t/2-1}
	\end{bmatrix} \;\in\reals^{t/2} \;,
\end{eqnarray}
for $d>2$ and for $d=2$,
$p^{(t)} = (2/t) {\mathbf 1}_{t/2}$ where ${\mathbf 1}_{t/2}$ is all ones vector in $\reals^{t/2}$.
There are $d(d-1)^{h-1}$ nodes at distance $h$ from the virtual source,
and by symmetry all of them are equally likely to have been the source:
\begin{eqnarray}
	\prob(G_T| v^*, \delta_H(v^*,v_t)=h) &=& \frac{1}{d(d-1)^{h-1}} p^{(t)}_{h} \nonumber\\
		&=& \frac{d-2}{d((d-1)^{t/2}-1)} \;,\nonumber
\end{eqnarray}
for $d>2$, which is independent of $h$.
Hence, all the infected nodes (except for the virtual source) are equally likely to have been the source of the origin. This statement is made precise in Equation \eqref{eq:p_diff}.

Together with the desired probability distribution in Equation \eqref{eq:MCp},
this gives a recursion over $t$ and $h$ for computing the appropriate $\alpha_d(t,h)$'s.
After some algebra and an initial state $p^{(2)}=1$, we get that the following choice ensures the desired
Equation \eqref{eq:MCp}:
\begin{eqnarray}
\alpha_d(t,h) = \left\{
  \begin{array}{lr}
    \frac{(d-1)^{t/2-h+1}-1}{(d-1)^{t/2+1}-1} & ~\text{if}~ d > 2\\
    \frac{t-2h+2}{t+2} & ~\text{if}~ d=2
  \end{array}
\right.
\label{eq:alpha}
\end{eqnarray}
With this choice of parameters, we show that adaptive diffusion spreads fast,
infecting $N_t = O((d-1)^{t/2})$ nodes at time $t$ and
each of the nodes except for the virtual source is equally likely to have been the source.

\begin{theorem}
\label{thm:main}
Suppose the contact network is a $d$-regular tree with $d\geq2$,
and one node $v^*$ in $G$ starts to spread a message according to
Protocol  \ref{alg:adp_diff} at time $t=0$.
At a certain time $T\geq0$ an adversary estimates the location of the source $v^*$
using the maximum likelihood estimator $\hv_{\rm ML}$.
The following properties hold for Protocol \ref{alg:adp_diff}:

\begin{itemize}
	\item [$(a)$] the number of infected nodes at time $T$ is
		\begin{eqnarray}
		N_{T}\geq \left\{
		  \begin{array}{lr}
			\frac{2(d-1)^{(T+1)/2}-d}{(d-2)}+1 & \text{if}~d>2\\
			T+1 & \text{if}~d=2
		\end{array}
		\right.
		\label{eq:n_diff}
		\end{eqnarray}
	\item [$(b)$] the probability of source detection for the maximum likelihood estimator at time $T$ is
	\begin{eqnarray}
		\prob\left(\hat{v}_{\rm ML}=v^*\right)  \leq \left\{
		  \begin{array}{lr}
		 \frac{d-2}{2(d-1)^{(T+1)/2}-d} &\text{if}~d>2\\
		(1/T) & \text{if}~d=2
		\label{eq:p_diff}
		\end{array}
		\right.
	\end{eqnarray}
	\item [$(c)$] the expected hop-distance between the true source $v^*$ and its estimate $\hat{v}_{\rm ML}$ under maximum likelihood estimation is lower bounded by
	\begin{eqnarray}
		\E[d(\hv_{\rm ML},v^*)] \geq
		\frac{d-1}{d}\frac{T}{2}.
		\label{eq:d_diff}
	\end{eqnarray}
\end{itemize}
\end{theorem}

Protocol \ref{alg:adp_diff} describes the details of the implementation of adaptive diffusion.
The first three steps are always the same.
At time $t=1$, the rumor source $v^*$ selects, uniformly at random, one of its neighbors to be the virtual source $v_2$ and passes the message to it.
Next at $t=2$, the new virtual source $v_2$ infects all its uninfected neighbors forming $G_2$ (see Figure \ref{fig:graph}).
Then node $v_2$ chooses to either keep the virtual source token with probability $\alpha_d(2,1)$ or to pass it along.

If $v_2$ chooses to remain the virtual source i.e., $v_4=v_2$, it passes `infection messages' to all the leaf nodes in the infected subtree, telling each leaf to infect all its uninfected neighbors. Since the virtual source is not connected to the leaf nodes in the infected subtree, these infection messages get relayed by the interior nodes of the subtree. This leads to $N_t$ messages getting passed in total (we assume this happens instantaneously). These messages cause the rumor to spread symmetrically in all directions at $t=3$. At $t=4$,
no more spreading occurs.

If $v_2$ does \emph{not} choose to remain the virtual source,
it passes the virtual source token to a randomly chosen neighbor $v_4$, excluding the previous virtual source (in this example, $v_0$).
Thus, if the virtual source moves, it moves away from the true source by one hop.
Once $v_4$ receives the virtual source token, it sends out infection messages. However, these messages do not get passed back in the direction of the previous virtual source. This causes the infection to spread asymmetrically over only one subtree of the infected graph ($G_3$ in left panel of Figure \ref{fig:graph}).
In the subsequent timestep ($t=4$), the virtual source remains fixed and passes the same infection messages again. After this second round of asymmetric spreading, the infected graph is once again symmetric about the virtual source $v_4$ ($G_4$ in left panel of Figure \ref{fig:graph}).

Although adaptive diffusion obfuscates the source on infinite regular trees, 
real world contact networks have cycles, and degrees are irregular, and the size is finite. 
We study how adaptive diffusion works under more realistic contact networks. 

\begin{algorithm}[ht!]
\caption{Adaptive Diffusion}
\label{alg:adp_diff}
\begin{algorithmic}[1]
\Require contact network $G=(V,E)$, source $v^*$, time $T$, degree $d$
\Ensure set of infected nodes $V_T$
\State $V_0 \gets \{v^*\}$, $h \gets 0$, $v_0 \gets v^*$
\State $v^*$ selects one of its neighbors $u$ at random
\State $V_1 \gets V_0 \cup \{u\}$, $h \gets 1$, $v_1 \gets u$
\State let $N(u)$ represent $u$'s neighbors
\State $V_2 \gets V_1 \cup N(u)\setminus \{v^*\}$, $v_2 \gets v_1$
\State $t \gets 3$
\For{$t \leq T$}
\State $v_{t-1}$ selects a random variable $X \sim U(0,1)$
\If{$X \leq \alpha_d(t-1, h)$}
\ForAll{$v \in N({v_{t-1}})$}
\State Infection Message({$G$,$v_{t-1}$,$v$,$G_t$})
\EndFor
\Else
\State $v_{t-1}$ randomly selects $u \in N({v_{t-1}})\setminus \{v_{t-2}\}$
\State $h \gets h + 1$
\State $v_{t} \gets u$
\ForAll{$v \in N({v_{t}})\setminus \left\{v_{t-1}\right\}$}
\State Infection Message({$G$,$v_{t}$,$v$,$V_t$})
\If{$t+1>T$}
\State \text{break}
\EndIf
\State Infection Message({$G$,$v_{t}$,$v$,$V_t$})
\EndFor
\EndIf
\State $t \gets t + 2$
\EndFor
\Procedure{Infection Message}{$G$,$u$,$v$,$V_t$}
\If{$v \in V_t$}
\ForAll{$w \in N({v})\setminus \left\{u\right\}$}
\State Infection Message({$G$,$v$,$w$,$G_t$})
\EndFor
\Else
\State $V_t \gets V_{t-2} \cup \{v\}$
\EndIf
\EndProcedure
\end{algorithmic}
\end{algorithm}

%% file: general_v4.tex
\section{General Contact Networks}
\label{sec:generalcn}
We study adaptive diffusion on general networks
 when the underlying graph is cyclic, irregular, and finite.

\subsection{Irregular tree networks}\label{sec:irregular_trees}
We first consider tree networks with potentially different degrees at the vertices.
Although the degrees are irregular,
we still apply adaptive diffusion
with $\alpha_{d_0}(t,h)$'s chosen for a specific $d_0$ that might be mismatched with the graph due to degree irregularities.
There are a few challenges in this degree-mismatched adaptive diffusion.
First, finding the maximum likelihood estimate of the source is not immediate, due to degree irregularities.
Second, it is not clear \emph{a priori} which choice of $d_0$ is good.
We first show an efficient message-passing algorithm for computing the maximum likelihood source estimate.
Using this estimate, we illustrate through simulations how  adaptive diffusion
performs and show that the detection probability is not too sensitive to the choice of $d_0$ as long as $d_0$ is above a threshold that depends on the degree distribution.

{\bf Efficient ML estimation}. To keep the discussion simple, we assume that $T$ is even. The same approach can be naturally extended to  odd $T$. Since the spreading pattern in adaptive diffusion is entirely deterministic given the sequence of virtual sources at each timestep,
computing the likelihood $\prob(G_T|v^*=v)$ is equivalent to computing the probability of the virtual source moving from $v$ to $v_T$ over $T$ timesteps. On trees, there is only one path from $v$ to $v_T$ and since we do not allow the virtual source to ``backtrack", we only need to compute the probability of every virtual source sequence $(v_0, v_2, \ldots, v_T)$ that meets the constraint  $v_0=v$. Due to the Markov property exhibited by adaptive diffusion, we have
$ \prob(G_T| \{(v_t, h_t)\}_{t\in\{2,4,\ldots,T\}}) = \prod_{\substack{t<T-1\\t\text{ even}}} \prob(v_{t+2}|v_{t}, h_t)$, where $h_t =\delta_H(v_0,v_t)$.
For $t$ even, $\prob(v_{t+2}|v_{t}, h_t)=\alpha_d(t, h_t)$ if $v_t=v_{t+2}$ and $\frac{1 - \alpha_d(t,h_t)}{\deg(v_{t})-1}$ otherwise.
Here $\deg(v_t)$ denotes the degree of node $v_t$ in $G$.
Given a virtual source trajectory $\mathcal P=(v_0, v_2, \ldots, v_T)$, let $\mathcal J_{\mathcal P}=(j_1, \ldots, j_{\delta_H(v_0,v_T)})$ denote the timesteps at which a new virtual source is introduced, with $1 \leq j_i \leq T$ . It always holds that $j_1=2$ because after $t=0$, the true source chooses a new virtual source and $v_2 \neq v_0$.
If the virtual source at $t=2$ were to keep the token exactly once after receiving it (so $v_2 = v_4$), then $j_2=6$, and so forth. To find the likelihood of a node being the true source, we sum over \emph{all} such trajectories
\begin{align}
\label{eq:likelihood}
&\prob(G_T|v_0) = \sum_{\mathcal J_{\mathcal P}: {\mathcal P}\in {\mathcal S(v_0,v_T,T)}} \underbrace{\frac{1}{\deg(v_0)}\prod_{k=1}^{\delta_H(v_0,v_T)-1} \frac{1}{\deg(v_{j_k})-1}}_{A_{v_0}} \times \nonumber\\
 & \underbrace{\prod_{\substack{t< T\\t \text{~even}}} \left(\mathbbm 1_{\{t+2 \notin \mathcal J_{\mathcal P}\}} \alpha_d(t,h_t) + {\mathbbm 1}_{\{t+2 \in \mathcal J_{\mathcal P}\}} (1-\alpha_d(t,h_t))\right), }_{B_{v_0}}
\end{align}
where $\mathbbm 1$ is the indicator function and $S(v_0,v_T,T)=\{{\mathcal P}: {\mathcal P}=(v_0,v_2,\ldots,v_T)\text{ is a valid trajectory of the virtual source}\}$.
Intuitively, part $A_{v_0}$ of the above expression is the probability of choosing the set of virtual sources specified by $\mathcal P$, and part $B_{v_0}$ is the probability of keeping or passing the virtual source token at the specified timesteps. Equation \eqref{eq:likelihood} holds for both regular and irregular trees.
Since the path between two nodes in a tree is unique, and part $A_{v_0}$
is (approximately) the product of node degrees in that path,
$A_{v_0}$ is identical
for all trajectories $\mathcal P$.
Pulling $A_{v_0}$ out of the summation, we wish to compute the summation over all valid paths $\mathcal P$ of part $B_{v_0}$
(for ease of exposition, we will use $B_{v_0}$ to refer to this whole summation).
Although there are combinatorially many valid paths, we can simplify the formula in Equation \eqref{eq:likelihood}
for the particular choice of $\alpha_d(t,h)$'s defined in \eqref{eq:alpha}.

\begin{propo}\label{prop:irregular}
Suppose that the underlying contact network $\tilde G$ is an infinite tree with degree of each node larger than one.
One node $\tilde v^*$ in $\tilde G$ starts to spread a message  at time $t=0$ according to
Protocol  \ref{alg:adp_diff} with the choice of $d=d_0$.
At a certain even time $T\geq 0$,
the maximum likelihood estimate of $\tilde v^*$ given a snapshot of the infected subtree $\tilde G_T$ is
\begin{eqnarray}
	\arg \max_{v\in \tilde G_T\setminus \tilde v_T} &&  \frac{d_0}{\deg(v)} 
\prod_{v' \in p(\tilde v_T,v)\setminus\{\tilde v_T,v\}} \frac{d_0-1}{ \deg(v')-1 }
		\;,\label{eq:mismatch}
\end{eqnarray}
where $\tilde v_T$ is the (Jordan) center of the infected subtree $\tilde G_T$,
$p(\tilde v_T, v)$ is the unique path from $\tilde v_T$ to $v$,
and $\deg(v')$ is the degree of node $v'$.
\end{propo}

\begin{figure}[h]
	    \centering
  \includegraphics[scale=0.64]{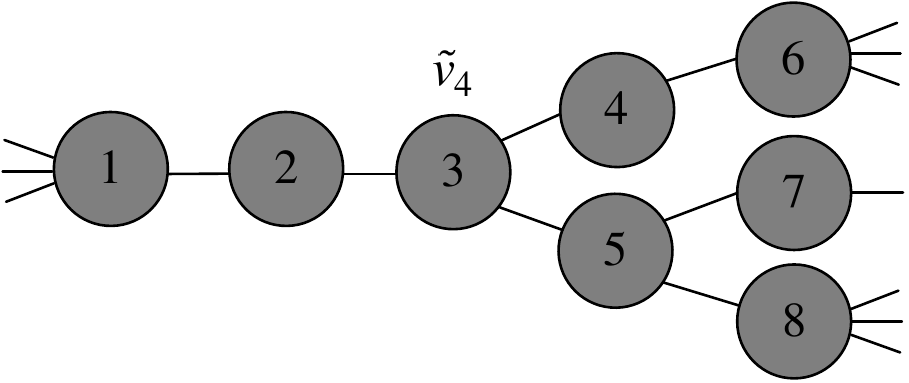}
  \caption{Irregular tree $\tilde G_4$ with virtual source $\tilde v_4$.}
  \label{fig:irregular}
\end{figure}

To understand this proposition, consider Figure \ref{fig:irregular}, which was spread using adaptive diffusion (Protocol \ref{alg:adp_diff}) with
 a choice of $d_0=2$.
  Then Equation \eqref{eq:mismatch} can be computed easily for each node, giving
  $[1/2,1,0,1,2/3,1/2,1/2,1/4]$ for nodes $[1,2,3,4,5,6,7,8]$, respectively. Hence, nodes 2 and 4 are most likely.
  Intuitively, nodes whose path to the center have small degrees are more likely.
  However, if we repeat this estimation assuming $d_0=4$, then
  Equation \eqref{eq:mismatch} gives $[3,2,0,2,4/3,3,3,3/2]$.
  In this case, nodes $1$, $6$, and $7$ are most likely.
  When $d_0$ is large, adaptive diffusion tends to place the source closer to the leaves of the infected subtree, so
  leaf nodes are more likely to have been the source.

\begin{proof}[of Proposition \ref{prop:irregular}]
We first make two observations:
($a$) Over regular trees, $\prob(G_T|u)=\prob(G_T|w)$ for any $u\neq w \in G_T$, even if they are different distances from the virtual souce. ($b$) Part $B_{v_0}$ is identical for regular and irregular graphs, as long as the distance from the candidate source node to $v_T$ is the same in both, and the same $d_0$ is used to compute $\alpha_{d_0}(t,h)$.
That is, let $\tilde G_T$ denote an infected subtree over an \emph{irregular} tree network, with virtual source $\tilde v_T$, and $G_T$ will denote a \emph{regular} infected subtree with virtual source $v_T$. For candidate sources $\tilde v_0 \in \tilde G_T$ and $v_0\in G_T$, if $\delta_H(\tilde v_T, \tilde v_0) = \delta_H(v_T, v_0)=h$, then $B_{v_0} = B_{\tilde v_0}$. So to find the likelihood of $\tilde v_0\in \tilde G_T$, we can solve for $B_{\tilde v_0}$ using the likelihood of $v_0\in G_T$, and compute $A_{\tilde v_0}$ using the degree information of every node in the infected, irregular subgraph.


To solve for $B_{\tilde v_0}$,
note that over regular graphs, $A_{v}=1/(d_0 \, (d_0-1)^{\delta_H(v,v_T)-1})$, where $d_0$ is the degree of the regular graph.
If $G$ is a regular tree,
Equation \eqref{eq:likelihood} still applies.
Critically, for regular trees,  the $\alpha_{d_0}(t,h)$'s are designed such that
the likelihood of each node being the true source is equal. Hence,
\begin{equation}
\prob(G_T|v_0) = \underbrace{\frac{1}{d_0(d_0-1)^{\delta_H(v_0,v_T)-1}}}_{A_{v_0}}  \, \times \,B_{v_0} \;,
\label{eq:leaf}
\end{equation}
is a constant that does not depend on $v_0$.
This gives $B_{v_0} \propto (d_0-1)^{\delta_H(v_T,v_0)}$.
%
%
From observation ($b$), we have that $B_{\tilde v_0} = B_{v_0}$.
Thus we get that for a $\tilde v_0 \in\tilde G_T\setminus \{\tilde v_T\}$,
\begin{eqnarray*}
	\prob(\tilde G_T |\tilde v_0) &=& A_{\tilde v_0} \, B_{\tilde v_0} \\
		&\propto& \frac{(d_0-1)^{\delta_H(\tilde v_T,\tilde v_0)}}{\deg(\tilde v_0) \prod_{\tilde v'\in p(\tilde v_T,\tilde v_0)\setminus \{\tilde v_0,\tilde v_T\}} (\deg(\tilde v')-1)}
\end{eqnarray*}
After scaling appropriately and noting that $|p(\tilde v_T,\tilde v_0)|=\delta_H(\tilde v_T, \tilde v_0)+1$,
this gives the formula in Equation \eqref{eq:mismatch}.
\end{proof}
%
\floatname{algorithm}{Algorithm}
\renewcommand{\algorithmicrequire}{\textbf{Input:}}
\renewcommand{\algorithmicensure}{\textbf{Output:}}
\begin{algorithm}
\caption{ML estimator of \eqref{eq:mismatch}}
\label{alg:ml_msg_passing}
\begin{algorithmic}[1]
\Require infected network $\tilde G_T=(\tilde V_T,\tilde E_T)$, virtual source $\tilde v_T$, time $T$, the spreading model parameter $d_0$
\Ensure $\text{argmax}_{\tilde v \in \tilde V_T}~\prob(\tilde G_T|\tilde v^*=\tilde v)$
\State $P_{\tilde v} \triangleq \prob(\tilde G_T|\tilde v^*=\tilde v)$.
\State $P_{\tilde v_T}\gets 0$
\State $A_{\tilde v} \gets 1$ for $\tilde v\in \tilde V_T \setminus \{\tilde v_T\}$
\State $A_{\tilde v_T}\gets 0$
\State $A\gets$ Degree Message($G_T$, $\tilde v_T$, $\tilde v_T$, $A$)
\State $\prob(G_T|v_{leaf}) \gets \frac{1}{d_0(d_0-1)^{T/2-1}}\prod _{\substack{t<T \\ t\text{ even}}}(1-\alpha_{d_0}(t,\frac{t}{2}))\}$
\ForAll {$\tilde v \in \tilde V_T \setminus \{\tilde v_T\}$}
\State $h\gets \delta_H(\tilde v, \tilde v_T)$
\State $B_{\tilde v}\gets \prob(G_T|v_{leaf})\cdot d_0 \cdot (d_0-1)^{h-1}$
\State $P_{\tilde v}\gets A_{\tilde v}\cdot B_{\tilde v}$
\EndFor
\Return argmax$_{\tilde v\in \tilde V_T}P_{\tilde v}$
\Procedure{Degree Message}{$\tilde G_T$, $\tilde u$, $\tilde v$, $A$}
\ForAll{$\tilde w \in N({\tilde v})\setminus \{\tilde u\}$}
\If{$\tilde v = \tilde u$}
\State $A_{\tilde w} \gets A_{\tilde v} / deg(\tilde w)$
\State Degree Message({$\tilde G_T$, $\tilde v$, $\tilde w$, $A$})
\Else
\If {$\tilde v$ is not a leaf}
\State $A_{\tilde w} \gets A_{\tilde v}\cdot deg(\tilde v) /(deg(\tilde w) \cdot (deg(\tilde v)-1))$
\State Degree Message({$\tilde G_T$, $\tilde v$, $\tilde w$, $A$})
\EndIf
\EndIf
\EndFor
\Return $A$
\EndProcedure
\end{algorithmic}
\end{algorithm}
\floatname{algorithm}{Protocol}
\renewcommand{\algorithmicrequire}{\textbf{Require}}
\renewcommand{\algorithmicensure}{\textbf{Ensure:}}
{\bf Implementation and numerical simulations.}
We provide an efficient
message passing algorithm for computing the
ML estimate in Equation \eqref{eq:mismatch}, which is naturally distributed.
We then use this estimator to simulate message spreading for random irregular trees and
show that when $d_0$ exceeds a threshold (determined by the degree distribution), obfuscation is not too sensitive to the choice of $d_0$.

$A_{\tilde v_0}$ can be computed efficiently for irregular graphs with a simple message-passing algorithm.
In this algorithm, each node $\tilde v$ multiplies its degree information by a cumulative likelihood that gets passed from the virtual source to the leaves. Thus if there are $\tilde N_T$ infected nodes in $\tilde G_T$, then $A_{\tilde v_0}$ for every $\tilde v_0 \in \tilde G_T$ can be computed by passing $O(\tilde N_T)$ messages. This message-passing is outlined in procedure `Degree Message' of Algorithm \ref{alg:ml_msg_passing}. For example, consider computing
$A_5$ for the graph in Figure \ref{fig:irregular}. The virtual source $\tilde v_T=3$ starts by setting $A_2=\frac{1}{2}$, $A_4=\frac{1}{2}$, and $A_5=\frac{1}{3}$. This gives $A_5$, but to compute other other values of $A_{\tilde w}$, the message passing continues. Each of the nodes $\tilde v\in N(3)$ in turn sets $A_{\tilde w}$ for \emph{their} children $\tilde w \in N(\tilde v)$; this is done by dividing $A_{\tilde v}$ by $deg(\tilde w)$ and replacing the factor of $\frac{1}{deg(\tilde v)}$ in $A_{\tilde v}$ with $\frac{1}{deg(\tilde v-1)}$. For example, node 5 would set $A_7=\frac{A_5}{2} \cdot \frac{3}{2}$.
This step is applied recursively until reaching the leaves.


As discussed earlier, $B_{\tilde v_0}$ only depends on $d_0$ and $\delta_H(\tilde v_T,\tilde v_0)$.
If $v_{\text{leaf}} \in G_T$ is a leaf node and $G$ is a regular tree, we get
\begin{equation}
\prob(G_T|v_{\text{leaf}})=\underbrace{\frac{1}{d_0(d_0-1)^{T/2-1}}}_{A_{v_\text{leaf}}}\underbrace{\prod_{\substack{t< T\\t\text{ even}}}(1-\alpha_{d_0}(t,\frac{t}{2})) }_{B_{v_\text{leaf}}}.
\label{eq:leaf}
\end{equation}
If $\tilde v_0$ is $h < T/2$ hops from $\tilde v_T$, then 
for node $v_0$ with $\delta_H(v_0,v_T)=h<T/2$ over a \emph{regular} tree,
\begin{eqnarray}\label{eq:b}
\prob(G_T|v_{0})= \prob(G_T|v_{\text{leaf}}) = \frac{1}{d_0\cdot(d_0-1)^{ h-1}} B_{v_0}. \nonumber
\end{eqnarray}
Finally, $B_{\tilde v_0} = B_{v_0}$. 
So to solve for $B_{5}$ in our example, we compute $\prob (G_T|v_{leaf})$ for a 3-regular graph at time $T=4$. This gives $\prob (G_4|v_{leaf})=A_{v_{leaf}}\cdot B_{v_{leaf}} = \frac{1}{6} \cdot (1-\alpha_3(2,1))=\frac{1}{9}$. Thus $B_5 = \prob (G_4|v_{leaf})\cdot d_0\cdot(d_0-1)^{h-1}=\prob (G_4|v_{leaf})\cdot 3\cdot(2)^{0}=\frac{1}{3}$. This gives
$\prob (\tilde G_4|5)=A_5\cdot B_5 = \frac{1}{9}$. The same can be done for other nodes in the graph to find the maximum likelihood source estimate.

We tested  adaptive diffusion  over random trees;  each node's degree was drawn i.i.d. from a fixed distribution.
\begin{figure}[h]
	    \centering
  \includegraphics[width=.44\textwidth]{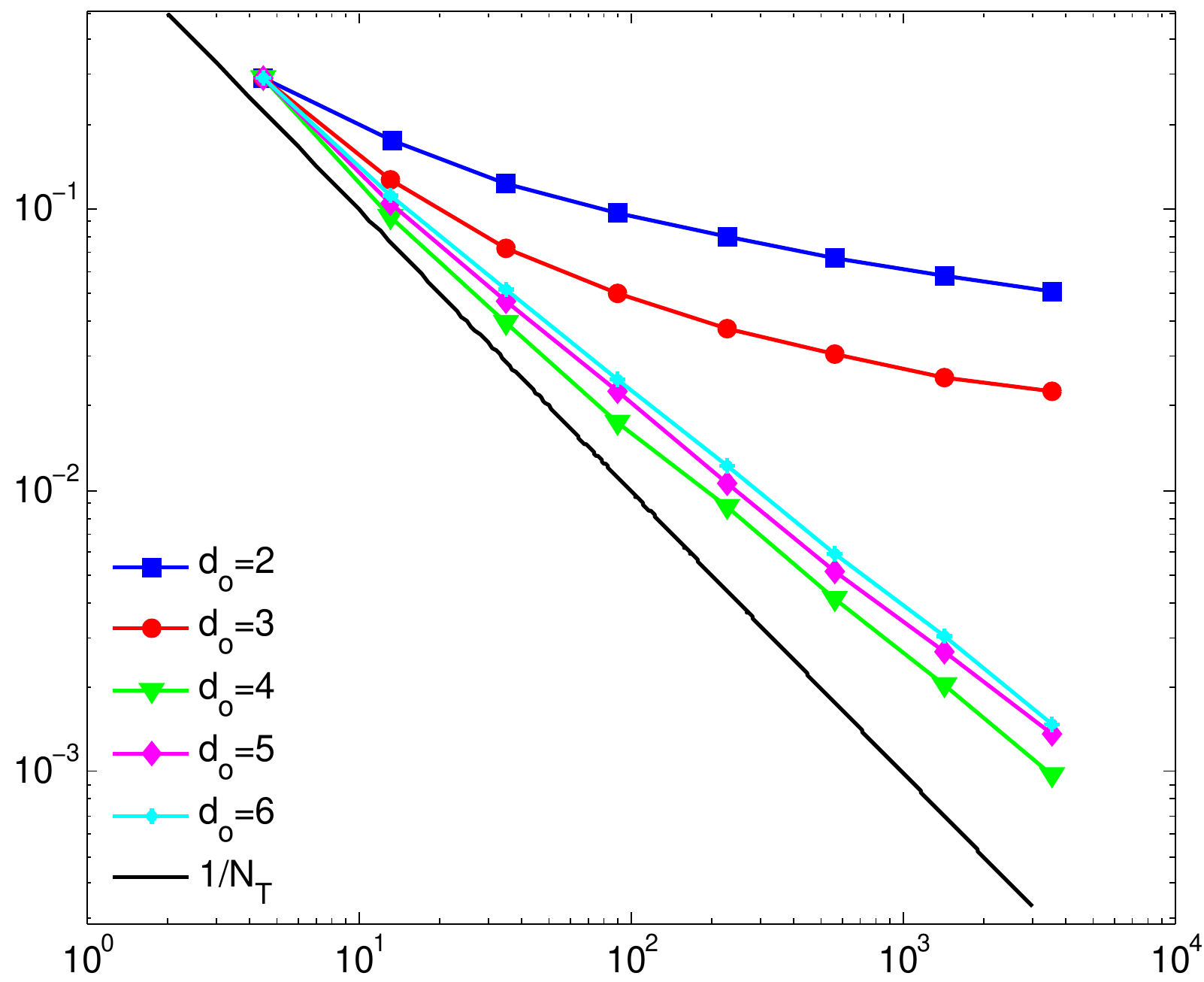}
  \put(-160,-10){Average size of infection, $N_T$}
	\put(-235,65){\rotatebox{90}{\large $\prob(\hv=v^*)$}}
  \caption{The probability of detection by the maximum likelihood estimator depends on the assumed degree $d_0$; the source cannot hide well below a threshold value of $d_0$.}
  \label{fig:irregular_adaptive}
\end{figure}

Figure \ref{fig:irregular_adaptive} illustrates simulation results for random trees in which each node has degree 3 or 4 with equal probability, averaged over 100,000 trials.
By the law of large numbers, the number of nodes infected scales as $N_T = e^{T\E[\log(D-1)]} \sim 2.45^T$, where $D$ represents the
degree distribution of the underlying random irregular tree.
The value of $d_0$ corresponds to a regular tree with size scaling as $(d_0-1)^T$.
Hence, one can expect that for $d_0-1 < 2.45$, the source is likely to be in the center of the infection, and
for $d_0>2.45$ the source is likely to be at the boundary of the infection.
Since the number of nodes in the boundary is significantly larger than the number of nodes in the center,
the detection probability is lower for $d_0-1>2.45$.
This is illustrated in the figure, which matches our prediction.
In general, $d_0=1+\lceil e^{\E[\log(D-1)]}\rceil$ provides the best obfuscation, and it is robust for any value above that.
In this plot, data points represent successive even timesteps; their uniform spacing implies the message is spreading exponentially quickly. 

\begin{figure}
	    \centering
  \includegraphics[width=.44\textwidth]{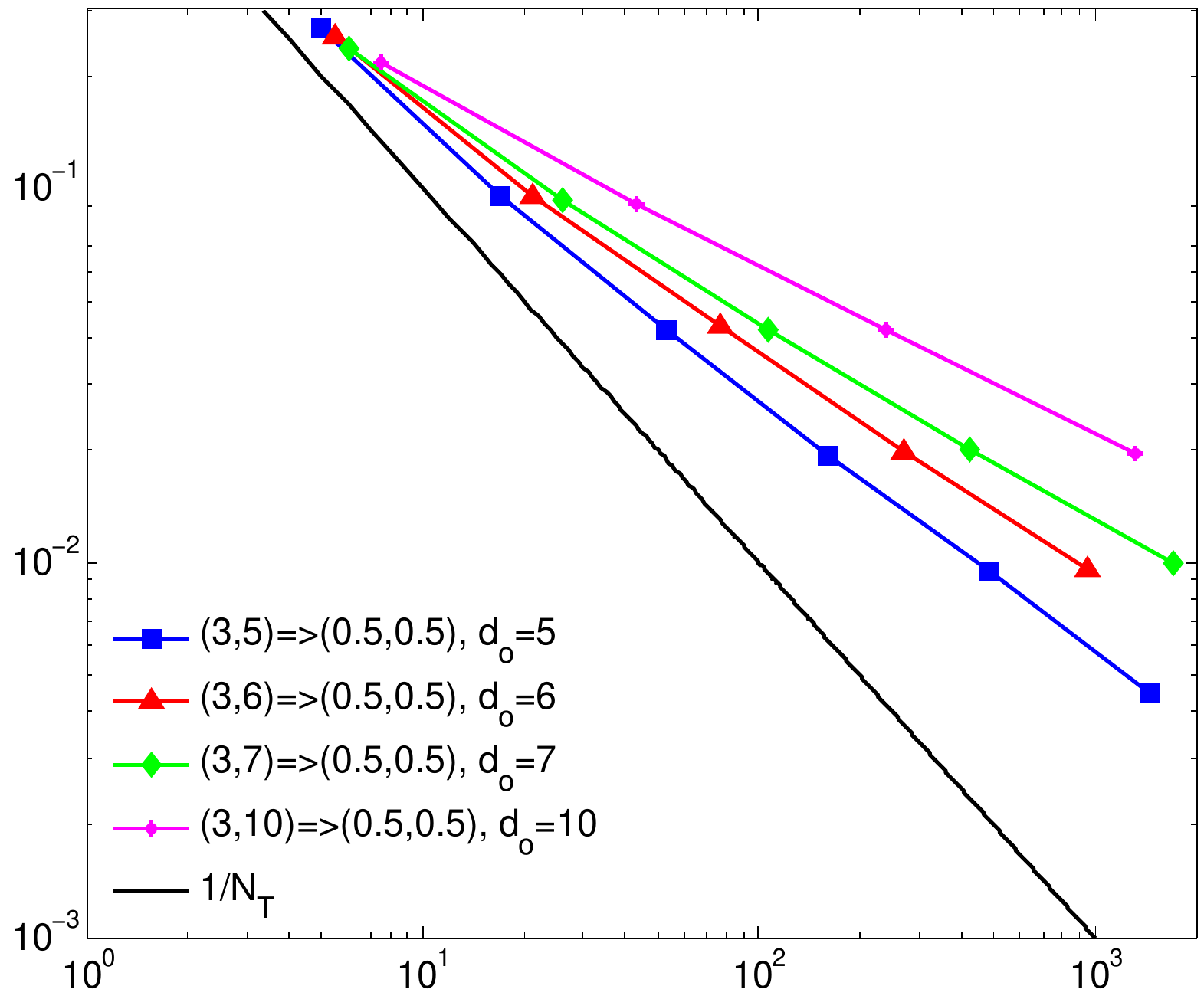}
  \put(-160,-10){Average size of infection, $N_T$}
	\put(-235,65){\rotatebox{90}{\large $\prob(\hv=v^*)$}}
  \caption{Adaptive diffusion no longer provides perfect obfuscation for highly irregular graphs.}
\label{fig:max_degree}
\end{figure}

Figure \ref{fig:max_degree} illustrates the probability of detection as a function of infection size while varying the degree distribution of the underlying tree. The notation $(3,5)=>(0.5,0.5)$ in the legend indicates that each node in the tree has degree 3 or 5, each with probability 0.5. For each distribution tested, we chose $d_0$ to be the maximum degree of each degree distribution.
The average size of infection scales as $N_T = e^{T\E[\log(D-1)]}$ as expected,
whereas the probability of detection scales as $(d_{\rm min}-1)^{-T} = 2^{-T}$, which is independent of the degree distribution.
This suggests that adaptive diffusion fails to provide near-perfect obfuscation when the underlying graph is irregular, and
the gap increases with the irregularity of the graph.

%

\subsection{Real World Networks}
\label{sec:numerical}
\begin{figure}[h]
	    \centering
  \includegraphics[width=.44\textwidth]{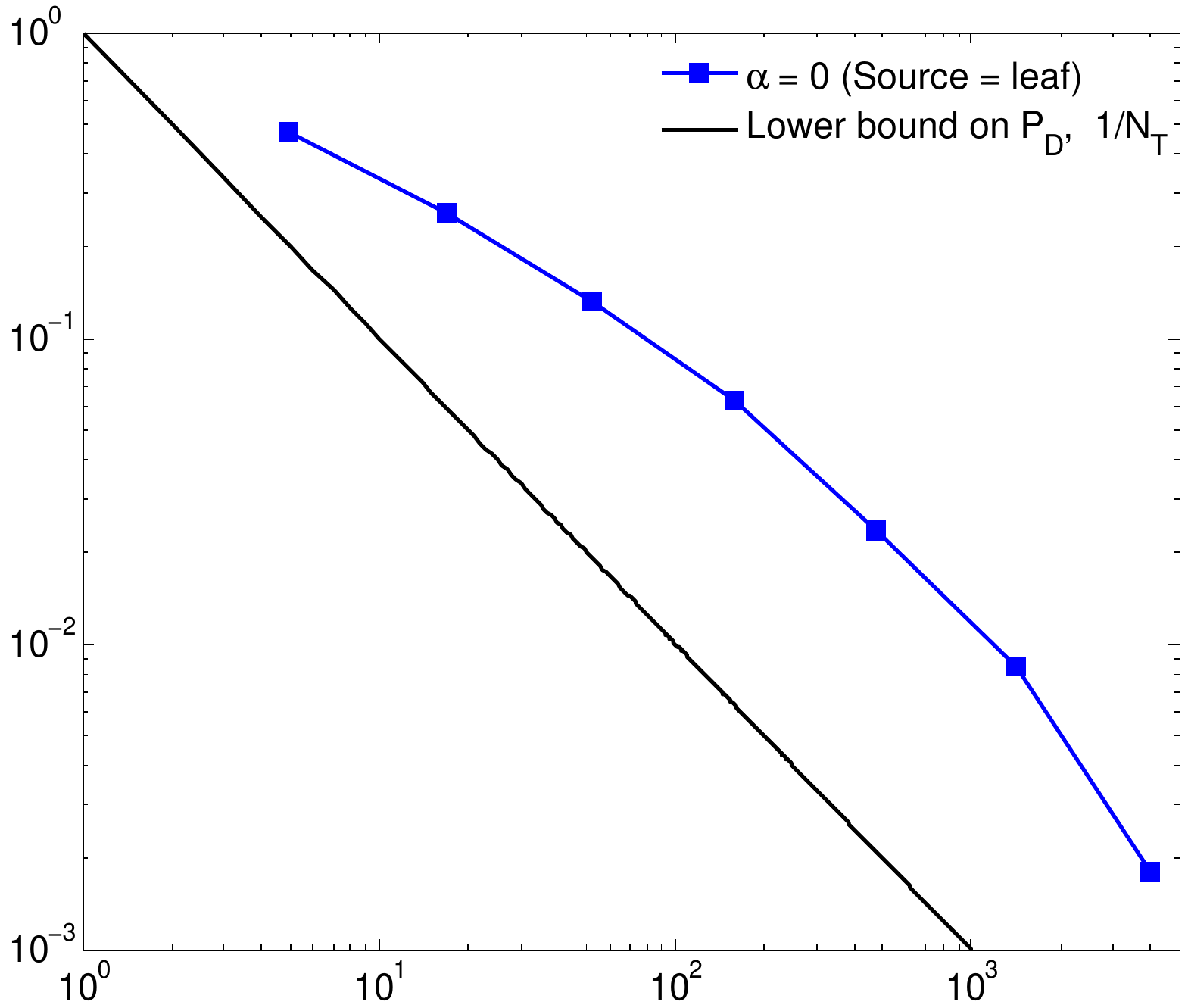}
  \put(-160,-10){Average size of infection, $N_T$}
	\put(-235,65){\rotatebox{90}{\large $\prob(\hv=v^*)$}}
  \caption{Near-ML probability of detection for the Facebook graph with adaptive diffusion.}
  \label{fig:facebook}
\end{figure}

To understand how adaptive diffusion fares in realistic scenarios that involve cycles, irregular degrees, and finite graph size,
we ran the adaptive diffusion protocol over an underlying connectivity network of 10,000 Facebook users in New Orleans circa 2009, as described by the Facebook WOSN dataset \cite{viswanath-2009-activity}.
We eliminated all nodes with fewer than three friends (this approach is taken by Secret so users cannot guess which of their friends originated the message), which left us with a network of 9,502 users.
Over this underlying network, we selected a node uniformly at random as the rumor source, and spread the message using adaptive diffusion setting with $d_0 = \infty$, which means that the virtual source is always passed to a new node.
This choice is to make the ML source estimation faster, and other choices of $d_0$ could outperform this naive choice.
To preserve the symmetry of our constructed trees as much as possible, we constrained each infected node to infect a maximum of three other nodes in each timestep.
We also give the adversary access to the undirected infection {\em subtree}
that explicitly identifies all pairs of nodes for which one node spread the infection to the other.
This subtree is overlaid on the underlying contact network, which is not necessarily tree-structured.
We demonstrate in simulation (Figure~\ref{fig:facebook})  that even with this strong side information, the adversary 
can only identify the true message source with low probability.

Using the naive method of enumerating every possible message trajectory, it is computationally expensive to find the exact ML source estimate since there are $2^T$ possible trajectories, depending on whether the virtual source stayed or moved  at each timestep. If the true source is one of the leaves, we can closely approximate the ML estimate \emph{among all leaf nodes}, using the same procedure as described in \ref{sec:irregular_trees}, with one small modification:
in  graphs with cycles, the term $(\text{deg}(v_{j_k})-1)$ from equation \eqref{eq:likelihood} should be substituted with $(\text{deg}_u(v_{j_k})-1)$, where $\text{deg}_u(v_{j_k})$ denotes the number of uninfected neighbors of $v_{j_k}$ at time $j_k$. Loops in the graph cause this value to be time-varying, and also dependent on the location of $v_0$, the candidate source.
We did not approximate the ML estimate for non-leaves because the simplifications used in Section \ref{sec:irregular_trees} to compute the likelihood no longer hold, leading to an exponential increase in the problem dimension.
This approach is only an approximation of the ML estimate because the virtual source could move in a loop over the social graph (i.e., the same node could be the virtual source more than once, in nonadjacent timesteps).

\begin{figure}[h]
	    \centering
  \includegraphics[width=.44\textwidth]{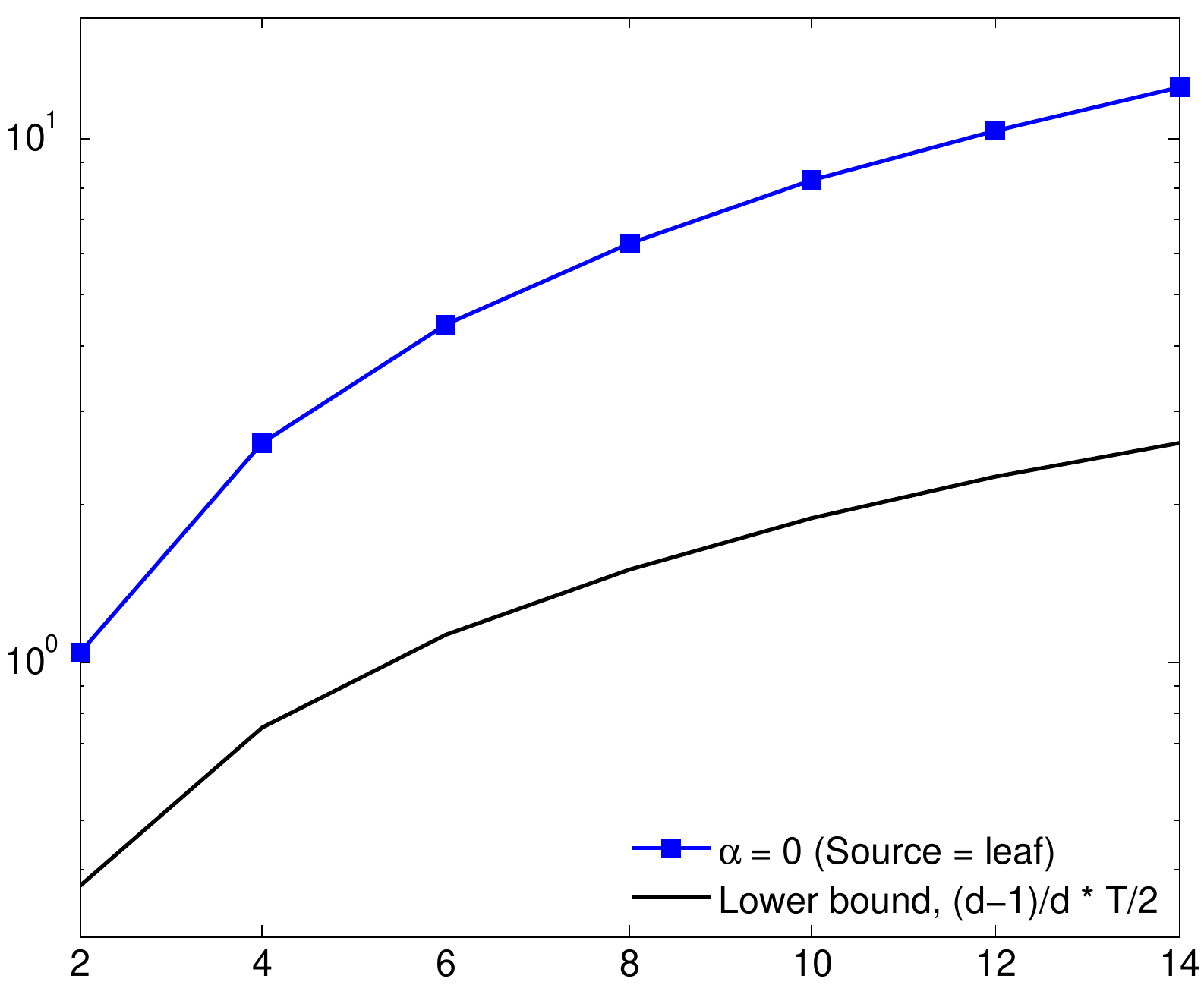}
  \put(-130,-10){Timestep, $T$}
	\put(-235,65){\rotatebox{90}{$\E[\delta_H(v^*,\hat{v}_{\rm ML})] $ }}
  \caption{ Hop distance between true source and estimated source over infection subtree for adaptive diffusion over the Facebook graph.}
  \label{fig:facebookhopdist}
\end{figure}

On average, adaptive diffusion reached 96 percent of the network within 10 timesteps using $d_0=4$.
We also computed the average distance of the true source from the estimated source \emph{over the infected subtree} (Figure \ref{fig:facebookhopdist}). We see that as time progresses, so does the hop distance of the estimated source from the true source.  In social networks, nearly everyone is within a small number of hops (say, 6 hops \cite{ugander2011anatomy}) from everyone else, so this computation is not as informative in this setting.
However, it is relevant in location-based connectivity graphs, which can induce large hop distances between nodes.

%% file: discussion.tex
\section{Discussion}
\label{sec:discussion}


Besides the adversarial model studied in this paper, anonymous messaging applications face challenges under alternative adversarial models that can occur in practice. 
Examples include $(a)$ an adversary that corrupts a subset of network nodes through malware, bribery, or Sybil node creation in order to access metadata like message timing and sender identity; 
$(b)$ an adversary that prevents nodes from following the messaging protocol (e.g., via malware); 
or $(c)$  an adversarial network provider that monitor all network activity and analyzes this activity retroactively. 

All these adversarial attacks increase the chance of source identification, which is a challenging problem for designing anonymous protocols.  
To a large extent, de-anonymization is an arms race in which there is always side information for an adversary to exploit. The point is to make that exploitation as expensive and difficult as possible, thereby preventing it from scaling. Within this arms race, anonymous spreading protocols ensure that adversaries cannot use message propagation patterns as a weapon.

%% file: protocol.tex
\section{Line and Tree Protocols}

\subsection{Line Protocol}
\begin{algorithm}[ht!]
\caption{Spreading on a line}
\label{alg:line}
\begin{algorithmic}[1]
\Require contact network $G=(V,E)$, source $v^*$, time $T$
\Ensure infected subgraph $G_T=(V_T,E_T)$
\State $V_0 \gets \{v^*\}$
\State $\delta_H(v^*-1,v^*) \gets 1$ and $\delta_H(v^*+1,v^*) \gets 1$
\State $t \gets 1$

\For{$t \leq T$}
\State $v \gets \text{ rightmost node in } V_t$
\State $\text{draw a random variable } X \sim U(0,1)$
\If{$X \leq  (\delta_H(v,v^*)+1)/(t+1)$}
\State $V_{t} \gets V_{t-1} \cup \{v+1\}$
\State  $\delta_H(v+1,v^*) \gets \delta_H(v,v^*) + 1$
\EndIf
\State $v \gets \text{ leftmost node in } V_t$
\State $\text{draw a random variable } Y \sim U(0,1)$
\If{$Y \leq  (\delta_H(v,v^*)+1)/(t+1)$}
\State $V_{t} \gets V_{t-1} \cup \{v-1\}$
\State  $\delta_H(v-1,v^*) \gets \delta_H(v,v^*) + 1$

\EndIf
\State $t \gets t + 1$
\EndFor
\end{algorithmic}
\end{algorithm}

Let the underlying contact network be an infinite line with $\V= \{..., -1, 0, 1, 2, ...\}$ and $E = \{(i,i+1) : i \in \mathbb{Z}\}$.
At time $t$, 
an infected node $v$ at the boundary spreads the infection to its uninfected neighbor with probability $p_{v,t}$ given in Equation \eqref{eq:linespread}.
`Infection' means transmitting the message, the hop-distance of the node (incremented by one), and the current timestep since the start of the outbreak (incremented by one).  Note that this protocol is naturally distributed.

\subsection{Tree Protocol}

\begin{algorithm}[h!]
\caption{Spreading on a tree}
\label{alg:tree}
\begin{algorithmic}[1]
\Require contact network $G=(V,E)$, source $v^*$, time $T$
\Ensure infected subgraph $G_T=(V_T,E_T)$

\State $V_0 \gets \{v^*\}$
\State $s_{1,v^*} \gets 0$ and $s_{2,v^*} \gets 0$
\State $v^*$ selects one of its neighbors $u$  at random
\State $V_1 \gets V_0 \cup \{u\}$
\State $s_{1,u} \gets 1$ and $s_{2,u} \gets 1$
\State $t \gets 2$

\For{$t \leq T$}
\ForAll{$v \in V_{t-1}$ with $s_{2,v} > 0$}
\If{$s_{1,v} = 1$}
\State $v$ selects one of its uninfected neighbors $u$ at random
\State $V_t \gets V_{t-1} \cup \{u\}$
\State $s_{1,u}\gets 1$ and $s_{2,u} \gets s_{2,v} + 1$
\State $s_{1,v}\gets 0$
\Else
\ForAll{uninfected neighboring nodes $w$ of $v$}
\State $V_t \gets V_{t-1} \cup \{w\}$
\State $s_{1,w}\gets 0$ and $s_{2,w} \gets s_{2,v} - 1$
\State $s_{2,v}\gets 0$
\EndFor
\EndIf
\EndFor
\State $t \gets t + 1$
\EndFor
\end{algorithmic}
\end{algorithm}

At time $t=0$, the source node $v^*$ is initialized with $s_{1,v^*}=1$ and $s_{2,v^*} = 0$.
At time $t=1$, $v^*$ selects, uniformly at random, one of its neighboring nodes (say node $u$) and passes the updated state variables $s_{1,u}= 1$ and $s_{2,u} = 1$ along with the message to $u$. The first state $s_{1,u}=1$ indicates that node $u$ is one of the `virtual sources'.
The second state $s_{2,u}=1$ indicates that node $u$ will eventually be at height $1$ in the infected subtree.
Afterwards, the protocol iterates the following procedures.
Any infected node $v$ that is not a future virtual source creates `offsprings'
that are not virtual sources $(s_{1,v'}=0)$ and
have heights decreased by one $(s_{1,v'}=s_{1,v}-1)$.
Any node $v$ with height zero (i.e., $s_{2,v}=0$) stops spreading the infection in order to ensure that $v$ and the true source are at the same depth from the center of $G_t$ for any $t$.
An infected node $v$ that is a future virtual source proceeds similarly, with one difference: it chooses one of its neighbors to be a virtual source with the second state incremented by one, i.e. $s_{2,v'}=s_{2,v}+1$.
Protocol \ref{alg:tree} summarizes this process and Figure \ref{fig:tree} illustrates that
if the contact network is a regular tree,
this process keeps the infected subtree balanced and regular, with the true source hiding among the leaves.

%

Figure \ref{fig:tree} shows an example of this spreading algorithm. At time $t=0$, node $v^*=0$ starts spreading a rumor in the network. At $t=1$, node $0$ infects node $1$ and sets $s_{1,1}= 1$ and $s_{2,1} = 1$. At $t=2$, node $1$ infects node $3$ and sets $s_{1,3}= 0$ and $s_{2,3} = 0$. Node $1$ also infects node $2$ and sets $s_{1,2}= 1$ and $s_{2,2} = 2$. At $t=3$, node $2$ infects node $4$ and sets $s_{1,4}= 0$ and $s_{2,4} = 1$. Node $2$ also infects node $5$ and sets $s_{1,5}= 1$ and $s_{2,5} = 3$. Observe that node $3$ will not infect any of its neighbors because $s_{1,3}= 0$ and $s_{2,3} = 0$.

by
sending a state information along with the message.
Each infected node $u$ keeps two state variables:
a binary variable  $s_{1,u}\in\{0,1\}$ and a non-negative integer $s_{2,u}\in\{0,1,2,\ldots,t\}$.
The binary state $s_{1,u}$ marks the future virtual sources of the infected subtree;
if $s_{1,u}=1$, then node $u$ will be the Jordan center of the graph at some point in the future.
The integer state $s_{2,u}$ keeps track of the eventual `height' of node $u$ in the infected subtree.
In the example above, node $3$ will always be at the leaf, so it has
$s_{2,3}=0$, where as node $9$ is currently a leaf, but eventually will be at
height $2$ and has $s_{2,9}=2$, since node $5$ will be at some point a center of
infected subtree of depth 3.
These state variables are computed for each node when it is infected, and
is constant over time.

%

%% file: proof.tex
\label{sec:proof}

\section{Proof of Proposition 2.1}
\label{apndx:proof1}
{\bf Spreading rate.} Consider the following time-dependent random walk on $\mathbb{N}_0=\{0,1, \ldots\}$. Let $x_t$ be the position of a black dot at time $t$ and assume that $x_0=0$. At time $t \geq 1$, the dot can either move to the right: $x_t = x_{t-1}+1$ with probability $p(t,x_{t-1})=(x_{t-1}+1)/(t+1)$; or stay where it is: $x_t = x_{t-1}$ with probability $q(t,x_{t-1})=1-p(t,x_{t-1})$. We claim that $\prob\left(x_T=k\right)=1/(T+1)$, for $k \in \{0,\ldots, T\}$. We will prove this by induction on $T$. Observe that for $T=1$, $x_1=x_0+1=1$ with probability $1/2$ and $x_1=0$ with probability $1/2$. Therefore, the claim is true for $T=1$. Assume it is true for $T-1$. At $t=T$,
\begin{eqnarray*}
\prob\left(x_T=0\right)&=&\prob\left(x_{T-1}=0\right)\left(1-\frac{1}{T+1}\right)=\frac{1}{T+1} \nonumber \\
\prob\left(x_T=T\right)&=&\prob\left(x_{T-1}=T\right)\frac{T}{T+1}=\frac{1}{T+1} \nonumber \\
\prob\left(x_T=k\right)&=&\prob\left(x_{T-1}=k-1\right)\frac{k}{T+1} + \nonumber \\
&& \prob\left(x_{T-1}=k\right)\left(1-\frac{k+1}{T+1}\right)=\frac{1}{T+1}, \nonumber \label{eq:lineconditional}
\end{eqnarray*}
for $k \in \{1, \ldots, T-1\}$. Thus, the claim is true for any $T$.

Given $v^*$ and $T$, consider an infected subgraph $G_T$ generated by Protocol \ref{alg:line} (Line Protocol). $G_T$ can be decomposed into two independent chains: the chain to the right of $v^*$ represented by $G^r_T$ and the chain to the left of $v^*$ represented by $G^l_T$. Therefore, we can write $G_T = G^r_T \cup G^l_T \cup \{v^*\}$, where $|G^r_T|,|G^l_T| \in \{0,\ldots, T\}$. The Line Protocol spreads on each side of $v^*$ according to the time-inhomogeneous random walk described above. Hence, $|G^r_T|$ and $|G^l_T|$ are independent and identically distributed, and $\prob\left(|G^r_T|=k\right)=1/(T+1)$ for any $k \in \{0,\ldots, T\}$. The number of infected nodes $N_T=|G_T|=|G^r_T|+|G^l_T|+1$ is the sum of two independent and uniformly distributed random variables. Thus, it is distributed according to the triangular distribution: 
\begin{equation}
\label{eq:n_line}
\prob\left(N_T=k\right)=\frac{1}{\left(T+1\right)^2}\left\{
\begin{array}{rl}
k, &  k \leq T+1,\\
	2(T+1) - k,   & k \leq 2T+1,\\
\end{array}\right.
\end{equation}

{\bf Probability of detection.}  Assume that the adversary observes an infected subgraph $G_T$ of size $N_T=|G_T|=k$. Moreover, assume that the adversary knows $T$. Not knowing $T$ will only lead to worse performance. If $k \leq T+1$, then for any $v \in G_T$ assumed to be the rumor source, we have that $|G^r_T|=k_r$, $|G^l_T|=k-k_r-1$ with $k_r \in \{0,1, \ldots, T\}$. Therefore, $\prob(G_T|v)=\prob(\left\{|G^r_T|=k_r\right\}\cap\left\{|G^l_T|=k-k_r-1\right\})=1/(T+1)^2$. This means that all $v \in G_T$ are equally likely and the maximum likelihood algorithm would select one node $v \in G_T$ at random. Hence, the probability of detecting the true source $\prob(\hv_{ML}=v^*|G_T,v^*)=1/(T+1)$ whenever $k \leq T+1$. If $k > T+1$, then for some nodes $v \in G_T$, we have that $|G^r_T|=k_r$, $|G^l_T|=k-k_r-1$ with $k_r>T$ or $|G^r_T|=k-k_r$, $|G^l_T|=k_l$ with $k_l>T$. These nodes could not have generated $G_T$ in $T$ timesteps and therefore $\prob(G_T|v)=0$ for all such nodes. The reader can verify that there are $2k-2T$ many such nodes. For the other nodes, one can easily verify (similar to above) that $\prob(G_T|v)=1/(T+1)^2$. Therefore, the probability of detecting the true source $\prob(\hv_{ML}=v^*|G_T,v^*)=1/(2T-k+1)$ whenever $k > T+1$. Putting it all together, we get that
\begin{eqnarray}
\prob(\hv_{ML}=v^*)&=&\sum_{G_T}\prob(\hv_{ML}=v^*|G_T)\prob(G_T|v^*) \nonumber \\
&=& \sum_{k=0}^{T}\frac{1}{k+1}\frac{k+1}{(T+1)^2}   \nonumber \\
&& ~~~~ + \sum_{T+1}^{2T}\frac{1}{2T-k+1}\frac{2T-k+1}{(T+1)^2} \nonumber \\
&=& \frac{2T+1}{(T+1)^2}.
\end{eqnarray}
Finally, observe that $\prob(\hv_{ML}=v^*)$ is not a function of $v^*$.


We can use the above analysis to compute the exact posterior distribution of the Line Protocol when the underlying contact network is a finite ring.
To put a prior on the source location, we fix the observation time $T$ and consider a finite ring of size larger than $2T+1$.
Assuming uniform prior for the source location on this finite ring,
the next remark provides
the posterior distribution of the source, given that we observe infected nodes $\{1,\ldots,m\}$ for some integer $m\leq 2T+1$.
\begin{remark}
	Let the infected nodes be $\{1,2,\ldots,m\}$, then
	\begin{eqnarray}
		\prob(v^*=k|G_T) &=& \frac{1}{Z_m} \sum_{t=\max\{m-k,k-1\}}^\infty \frac{1}{(t+1)^2} \left(\frac{m+1}{(t+2)}\right. \nonumber \\
&& ~~~~~~~~~~~~~~~~~-\left.\frac{k(m-k+1)}{(t+2)^2} \right)\;,
		\label{eq:lineposterior}
	\end{eqnarray}
	for all $k\in[m]$
	where $Z_m$ is the normalizing constant to ensure the probabilities sum to one.
	\label{rem:lineposterior}
\end{remark}

\begin{proof}[of Remark \ref{rem:lineposterior}]
	Denote the $m$ infected nodes by $\{1,\ldots,m\}$,
	and let $G^*$ denote the infected subgraph. Then, for $k\leq (1+m)/2$,
	\begin{align}
		&\prob(\text{there exists a time $T$ such that }G_T = G^*|v^*=k) \\
		&= \sum_{t=m-k}^{\infty} \prob(G_t = G^*\text{ and } G_{t+1}\neq G^*| v^*=k) \\
		&= \sum_{t=m-k}^{\infty} \frac{1}{(1+t)^2}\Big(1- \Big(\frac{t+2-k}{t+2}\Big)\Big(\frac{t+2-(m-k+1)}{t+2} \Big)\,\Big)\;,
	\end{align}
	where the last line follows from Equation \eqref{eq:lineconditional},
	where $\prob(|G^l_T|=k-1)= 1/(T+1)$, $\prob(|G^r_T|=m-k)= 1/(T+1)$,
	and $ \prob(\text{no infection at time $t+1$}|G_t=G^*\text{ and }v^*=k)=((t+2-k)(t+1+k-m)/(t+2)^2)$.
	For $k\geq(1+m)/2$, we have a similar formula with summation starting from $k-1$.
	With the uniform prior on the finite ring, this proves Equation \eqref{eq:lineposterior}.
\end{proof}

\section{Proof of Proposition 2.2}
\label{apndx:proof2}
First, under Protocol \ref{alg:tree} (Tree Protocol), $G_T$ is a complete $(d-1)$-ary tree (with the exception that the root has $d$ children) of depth $T/2$ whenever $T$ is even. $G_T$ is made up of two complete $(d-1)$-ary trees of depth $(T-1)/2$ each with their roots connected by an edge whenever $T$ is odd. Therefore, it follows that $N_T$ is a deterministic function of $T$ and is given by
	\begin{equation}
	\label{eq:n_tree2}
	N_T \;=\;
	\left\{
	\begin{array}{rl}
	1, &  T=0,\\
	\frac{2(d-1)^{(T+1)/2}}{d-2}-\frac{2}{d-2},   & T \geq 1, ~T\text{ odd}\;, \\
	\frac{d(d-1)^{T/2}}{d-2}-\frac{2}{d-2},   & T \geq 2, ~T\text{ even}\;;
	\end{array}\right.
	\end{equation}
 The lower bound on $N_T$ in Equation \eqref{eq:n_tree} follows immediately from the above expression.

 For any given infected graph $G_T$, it can be verified that any non-leaf node could not have generated $G_T$ under the Tree Protocol. In other words, $\prob(G_T|v \text{ non-leaf node}) =0$ and $v$ could not have started the rumor. On the other hand, we claim that for any two leaf nodes $v_1,v_2 \in G_T$, we have that $\prob(G_T|v_1)= \prob(G_T|v_2) >0$. This is true because for each leaf node $v \in G_T$, there exists a sequence of state values $\left\{s_{1,u},s_{2,u}\right\}_{u \in G_T}$ that evolves according to the Tree Protocol with $s_{1,v}= 1$ and $s_{2,v}=0$. Further, the regularity of the underlying graph $G$ ensures that all these sequences are equally likely. Therefore, the probability of correct rumor source detection under the maximum likelihood algorithm is given by $\prob_{ML}(T) = 1/N_{l,T}$, where $N_{l,T}$ represents the number of leaf nodes in $G_T$. It can be also shown that $N_{l,T}$ and $N_T$ are related to each other by the following expression
\begin{equation}
N_{l,T} = \frac{(d-2)N_T+2}{d-1}.
\end{equation}
This proves expression for $\prob\big( \hv_{\rm ML} =v^* \big)$ given in \eqref{eq:pd_tree}.

{\bf Expected distance.}
For any $v^* \in G$ and any $T$, $\E[\delta_H(v^*,\hv_{\rm ML})]$ is given by
\begin{equation}
\label{eq:exp_d}
\E[\delta_H(v^*,\hv_{\rm ML})]= \sum_{v \in G}\sum_{G_T}\prob(G_T|v^*)\prob(\hv_{\rm ML} =v)\delta_H(v^*,v).
\end{equation}
As indicated above, no matter where the rumor starts from, $G_T$ is a $(d-1)$-ary tree (with the exception that the root has $d$ children) of depth $T/2$ whenever $T$ is even. Moreover, $\hv_{\rm ML}=v$ with probability $1/N_{l,T}$ for all $v$ leaf nodes in $G_T$. Therefore, the above equation can be solved exactly to obtain the expression provided in the statement of the proposition.
\section{Proof of Theorem 3.1}
\balance
{\bf Spreading rate.} Once again, under Protocol \ref{alg:adp_diff}, $G_T$ is a complete $(d-1)$-ary tree (with the exception that the root has $d$ children) of depth $T/2$ whenever $T$ is even. Whenever $T$ is odd, with probability $\alpha(T,h)$, $G_T$ is again such a $(d-1)$-ary tree of depth $(T+1)/2$. With probability $1-\alpha(T,h)$, $G_T$ is made up of two $(d-1)$-ary trees of depth $(T-1)/2$ each with their roots connected by an edge. Therefore, it follows that when $d>2$, $N_T$ is given by
	\begin{equation}
	\label{eq:n_tree2}
	N_T \;=\;
	\left\{
	\begin{array}{rl}
	1, &  T=0,\\
	\frac{2(d-1)^{(T+1)/2}}{d-2}-\frac{2}{d-2},   & T \geq 1, ~T\text{ odd, w.p. }(1-\alpha)\;, \\
	\frac{d(d-1)^{(T+1)/2}}{d-2}-\frac{2}{d-2},   & T \geq 1, ~T\text{ odd, w.p. }\alpha\;, \\
	\frac{d(d-1)^{T/2}}{d-2}-\frac{2}{d-2},   & T \geq 2, ~T\text{ even}\;;
	\end{array}\right.
	\end{equation}
Similarly, when $d=2$, $N_T$ can be expressed as follows:
	\begin{equation}
	\label{eq:n_tree2_d2}
	N_T \;=\;
	\left\{
	\begin{array}{rl}
	1, &  T=0,\\
	T+1,   & T \geq 1, ~T\text{ odd, w.p. }(1-\alpha)\;, \\
	T+2,   & T \geq 1, ~T\text{ odd, w.p. }\alpha\;, \\
	T+2,   & T \geq 2, ~T\text{ even}\;;
	\end{array}\right.
	\end{equation}
 The lower bound on $N_T$ in Equation \eqref{eq:n_tree} follows immediately from the above expressions.

{\bf Probability of detection.} For any given infected graph $G_T$, the virtual source $v_T$ cannot have been the source node, since the true source always passes the token at timestep $t=1$. So $\prob (G_T|v=v_T)=0$. We claim that for any two nodes that are not the virtual source at time $T$, $u, w\in G_T$, $\prob(G_T|u)=\prob(G_T|w)>0$. This is true iff for any non-virtual-source node $v$, there exists a sequence of virtual sources ${v_i}_{i=0}^T$ that evolves according to Protocol \ref{alg:adp_diff} with $v_0=v$ that results in the observed $G_T$, and for all $u,w\in G_T\setminus \{v_T\}$, this sequence has the same likelihood. 
In a tree, a unique path exists between any pair of nodes, so we can always find a valid path of virtual sources from a candidate node $u \in G_T \setminus \{v_T\}$ to $v_T$. We claim that any such path leads to the formation of the observed $G_T$. 
Due to regularity of $G$ and the symmetry in $G_T$, for even $T$, $\prob(G_T| v^{(1)}) =  \prob(G_T| v^{(2)})$  for all $v^{(1)},v^{(2)} \in G_T$ with $\delta_H(v^{(1)},v_T) = \delta_H(v^{(2)},v_T)$. Moreover, recall that the $\alpha_d(t,h)$'s were designed to satisfy the distribution in Equation \ref{eq:MCp}. Combining these two observations with the fact that we have $(d-1)^h$ infected nodes $h$-hops away from the virtual source, we get that for all $v^{(1)}, v^{(2)} \in G_T\setminus \{ v_T\}$, $\prob(G_T| v^{(1)}) = \prob(G_T| v^{(2)})$. For odd $T$, if the virtual source remains the virtual source, then $G_T$ stays symmetric about $v_T$, in which case the same result holds. If the virtual source passes the token, then $G_T$ is perfectly symmetric about the edge connecting $v_{T-1}$ and $v_T$. Since both nodes are virtual sources (former and present, respectively) and $T>1$, the adversary can infer that neither node was the true source. Since the two connected subtrees are symmetric and each node within a subtree has the same likelihood of being the source by construction (Equation \ref{eq:MCp}), we get that for all $v^{(1)}, v^{(2)} \in G_T\setminus \{v_T,v_{T-1}\}$, $\prob(G_T| v^{(1)}) = \prob(G_T| v^{(2)})$. Thus at odd timesteps, $\prob(\hat v_{ML}=v^*)\geq 1/(N_T-2)$.

%% file: sig2015_R1.bbl
\begin{thebibliography}{10}

\bibitem{secret}
Secret.
\newblock \url{https://www.secret.ly}.

\bibitem{whisper}
Whisper.
\newblock \url{http://whisper.sh}.

\bibitem{yikyak}
Yik yak.
\newblock \url{http://www.yikyakapp.com/}.

\bibitem{chaum88}
D.~Chaum.
\newblock The dining cryptographers problem: Unconditional sender and recipient
  untraceability.
\newblock {\em Journal of cryptology}, 1(1), 1988.

\bibitem{freenet}
I.~Clarke, O.~Sandberg, B.~Wiley, and T.W. Hong.
\newblock Freenet: A distributed anonymous information storage and retrieval
  system.
\newblock In {\em Designing Privacy Enhancing Technologies}. Springer, 2001.

\bibitem{corrigan2010dissent}
H.~Corrigan-Gibbs and B.~Ford.
\newblock Dissent: accountable anonymous group messaging.
\newblock In {\em Proc. CCS}. ACM, 2010.

\bibitem{freeHavenProject}
R.~Dingledine, M.J. Freedman, and D.~Molnar.
\newblock The free haven project: Distributed anonymous storage service.
\newblock In {\em Designing Privacy Enhancing Technologies}. Springer, 2001.

\bibitem{tor}
R.~Dingledine, N.~Mathewson, and P.~Syverson.
\newblock Tor: The second-generation onion router.
\newblock Technical report, DTIC Document, 2004.

\bibitem{FC12}
V.~Fioriti and M.~Chinnici.
\newblock Predicting the sources of an outbreak with a spectral technique.
\newblock {\em arXiv preprint arXiv:1211.2333}, 2012.

\bibitem{tarzan}
M.J. Freedman and R.~Morris.
\newblock Tarzan: A peer-to-peer anonymizing network layer.
\newblock In {\em Proc. CCS}. ACM, 2002.

\bibitem{goel2003herbivore}
S.~Goel, M.~Robson, M.~Polte, and E.~Sirer.
\newblock Herbivore: A scalable and efficient protocol for anonymous
  communication.
\newblock Technical report, Cornell University, 2003.

\bibitem{golle2004dining}
P.~Golle and A.~Juels.
\newblock Dining cryptographers revisited.
\newblock In {\em Advances in Cryptology-Eurocrypt 2004}. Springer, 2004.

\bibitem{luo2013identify}
Wuqiong Luo, W~Tay, and Mei Leng.
\newblock How to identify an infection source with limited observations.
\newblock 2013.

\bibitem{Austin3}
E.~A. Meirom, C.~Milling, C.~Caramanis, S.~Mannor, A.~Orda, and S.~Shakkottai.
\newblock Localized epidemic detection in networks with overwhelming noise.
\newblock 2014.

\bibitem{Austin4}
C.~Milling, C.~Caramanis, S.~Mannor, and S.~Shakkottai.
\newblock On identifying the causative network of an epidemic.
\newblock In {\em Allerton Conference}, pages 909--914, 2012.

\bibitem{Austin2}
C.~Milling, C.~Caramanis, S.~Mannor, and S.~Shakkottai.
\newblock Detecting epidemics using highly noisy data.
\newblock In {\em MobiHoc}, pages 177--186, 2013.

\bibitem{Austin1}
Chris Milling, Constantine Caramanis, Shie Mannor, and Sanjay Shakkottai.
\newblock Network forensics: Random infection vs spreading epidemic.
\newblock In {\em Proceedings of the 12th ACM SIGMETRICS/PERFORMANCE Joint
  International Conference on Measurement and Modeling of Computer Systems},
  SIGMETRICS '12, pages 223--234, New York, NY, USA, 2012. ACM.

\bibitem{PVF12}
B.~A. Prakash, J.~Vreeken, and C.~Faloutsos.
\newblock Spotting culprits in epidemics: How many and which ones?
\newblock In {\em ICDM}, volume~12, pages 11--20, 2012.

\bibitem{SZ11b}
D.~Shah and T.~Zaman.
\newblock Finding rumor sources on random graphs.
\newblock {\em arXiv preprint arXiv:1110.6230}, 2011.

\bibitem{SZ11a}
D.~Shah and T.~Zaman.
\newblock Rumors in a network: Who's the culprit?
\newblock {\em Information Theory, IEEE Transactions on}, 57(8):5163--5181, Aug
  2011.

\bibitem{ugander2011anatomy}
Johan Ugander, Brian Karrer, Lars Backstrom, and Cameron Marlow.
\newblock The anatomy of the facebook social graph.
\newblock {\em arXiv preprint arXiv:1111.4503}, 2011.

\bibitem{viswanath-2009-activity}
Bimal Viswanath, Alan Mislove, Meeyoung Cha, and Krishna~P. Gummadi.
\newblock On the evolution of user interaction in facebook.
\newblock In {\em Proceedings of the 2nd ACM SIGCOMM Workshop on Social
  Networks (WOSN'09)}, August 2009.

\bibitem{von2003k}
L.~von Ahn, A.~Bortz, and N.J. Hopper.
\newblock K-anonymous message transmission.
\newblock In {\em Proc. CCS}. ACM, 2003.

\bibitem{WDZT14}
Zhaoxu Wang, Wenxiang Dong, Wenyi Zhang, and Chee~Wei Tan.
\newblock Rumor source detection with multiple observations: Fundamental limits
  and algorithms.
\newblock 2014.

\bibitem{ZY13}
K.~Zhu and L.~Ying.
\newblock A robust information source estimator with sparse observations.
\newblock {\em arXiv preprint arXiv:1309.4846}, 2013.

\end{thebibliography}
